 \documentclass{emulateapj} 
 \usepackage{lscape}
 
\newcommand{\etal}{\mbox{et~al.}}

\def\deg      {{\ifmmode^\circ\else$^\circ$\fi}} 

 \shorttitle{AGN Morphologies}
 \shortauthors{Gabor et al.}
 
\begin{document}
 
 
 \title{AGN Host Galaxy Morphologies in COSMOS \altaffilmark{$\star$}}
 

%
%
 \author{ 
J. M. Gabor\altaffilmark{1},
C. D. Impey\altaffilmark{1},
K. Jahnke\altaffilmark{2},
B. D. Simmons\altaffilmark{3},
J. R. Trump\altaffilmark{1},
A. M. Koekemoer\altaffilmark{4},
M. Brusa\altaffilmark{5},
N. Cappelluti\altaffilmark{5},
E. Schinnerer\altaffilmark{2},
V. Smol\v{c}i\'{c}\altaffilmark{6},
M. Salvato\altaffilmark{6},
J. D. Rhodes\altaffilmark{6,7},
B. Mobasher\altaffilmark{8},
P. Capak\altaffilmark{6},
R. Massey\altaffilmark{6},
A. Leauthaud\altaffilmark{9},
N. Scoville\altaffilmark{6}
}
 
 
\altaffiltext{$\star$}{Based on observations with the NASA/ESA {\em
Hubble Space Telescope}, obtained at the Space Telescope Science
Institute, which is operated by AURA Inc, under NASA contract NAS
5-26555; also based on data collected at : the Subaru Telescope, which is operated by
the National Astronomical Observatory of Japan; the XMM-Newton, an ESA science mission with
instruments and contributions directly funded by ESA Member States and NASA; the European Southern Observatory under Large Program 175.A-0839, Chile; Kitt Peak National Observatory, Cerro Tololo Inter-American
Observatory, and the National Optical Astronomy Observatory, which are
operated by the Association of Universities for Research in Astronomy, Inc.
(AURA) under cooperative agreement with the National Science Foundation; 
the National Radio Astronomy Observatory which is a facility of the National Science 
Foundation operated under cooperative agreement by Associated Universities, Inc ; 
and and the Canada-France-Hawaii Telescope with MegaPrime/MegaCam operated as a
joint project by the CFHT Corporation, CEA/DAPNIA, the National Research
Council of Canada, the Canadian Astronomy Data Centre, the Centre National
de la Recherche Scientifique de France, TERAPIX and the University of
Hawaii.}  
\altaffiltext{1}{Steward Observatory, University of Arizona, 933 North Cherry Avenue, Tucson, AZ 85721; jgabor@as.arizona.edu}
\altaffiltext{2}{Max-Planck-Institut f\"ur Astronomie, K\"onigstuhl 17, Heidelberg, D-69117, Germany}
\altaffiltext{3}{Department of Astronomy, Yale University, P.O. Box 208101, New Haven, CT 06520-8101}
\altaffiltext{4}{Space Telescope Science Institute, 3700 San Martin Drive, Baltimore, MD 21218}
\altaffiltext{5}{Max-Planck-Institut f\"ur extraterrestrische Physik, Giessenbachstrasse 1, D-85478 Garching, Germany}
\altaffiltext{6}{California Institute of Technology, MC 105-24, 1200 East California Boulevard, Pasadena, CA 91125}
\altaffiltext{7}{Jet Propulsion Laboratory, California Institute of Technology, Pasadena, CA 91109}
\altaffiltext{8}{Physics and Astronomy Department, University of California, Riverside, Ca 92521}
\altaffiltext{9}{BNL \& BCCP, University of California, Berkeley, CA 94720}
%

  
\begin{abstract}
We use HST/ACS images and a photometric catalog of the COSMOS field to analyze morphologies of the host galaxies of $\sim$400 AGN candidates at redshifts $0.3 < z < 1.0$.  We compare the AGN hosts with a sample of non-active galaxies drawn from the COSMOS field to match the magnitude and redshift distribution of the AGN hosts.  We perform 2-D surface brightness modeling with GALFIT to yield host galaxy and nuclear point source magnitudes.  X-ray selected AGN host galaxy morphologies span a substantial range that peaks between those of early-type, bulge-dominated and late-type, disk-dominated systems.  We also measure the asymmetry and concentration of the host galaxies.  Unaccounted for, the nuclear point source can significantly bias results of these measured structural parameters, so we subtract the best-fit point source component to obtain images of the underlying host galaxies.  Our concentration measurements reinforce the findings of our 2-D morphology fits, placing X-ray AGN hosts between early- and late-type inactive galaxies.  AGN host asymmetry distributions are consistent with those of control galaxies.  Combined with a lack of excess companion galaxies around AGN, the asymmetry distributions indicate that strong interactions are no more prevalent among AGN than normal galaxies.  In light of recent work, these results suggest that the host galaxies of AGN at these X-ray luminosities may be in a transition from disk-dominated to bulge-dominated, but that this transition is not typically triggered by major mergers.
\end{abstract}

 
 

 

\section{Introduction}
\label{sec.intro}
 
Studies over the past decade suggest fundamental links between galaxies and their central supermassive black holes (SMBHs).  The masses of the central SMBHs in nearby galaxies correlate with several host bulge properties, including luminosity \citep{kor95,mcl01,mar03}, velocity dispersion \citep{fer00, geb00}, and mass \citep[for a review of these relations, see Ferrarese 2004]{mag98}.  More recent studies extend these relationships to galaxies and quasars at redshifts from $z=0.37$ \citep{tre04} to as high as $z\sim4$ \citep{pen06a, pen06b}.

These observations indicate co-evolution of SMBH mass accretion and host bulge formation processes, perhaps through interactions such as AGN feedback.
Galaxy merger simulations that include a prescription for SMBH feedback \citep{spr05} and reproduce the $M_{BH}$-$\sigma_{bulge}$ relation can recover several observed properties of quasar host descendants, including the stellar mass function of local elliptical galaxies, and the red galaxy luminosity function and its variation with redshift \citep{hop06}.  In these models, gas-rich mergers drive material toward the central black holes, leading to intense star formation and SMBH accretion.  The nuclear SMBH begins its active phase obscured, but feedback energy during the peak accretion phase blows away the obscuring material and results in a brief quasar phase.  The blowout phase coincides with a rapid truncation of star formation throughout the host galaxy.  Recent observational studies lend credibility to this picture by hinting that star-formation quenching coincides with AGN activity \citep{bun07, sil08, tre07}.

This picture of SMBH-host co-evolution relies on a hypothesized merger mechanism for fueling active black holes, and incorporates predictions that gas-rich disk galaxies merge to form luminous starbursts, eventually evolving into massive elliptical galaxies.  Other possible fueling mechanisms include less-direct tidal interactions with nearby galaxies \citep{men04} and instabilities in a quiescent galaxy's gaseous disk \citep[e.g.][]{hop06b}.  In this study, we explore the interaction mechanism for intermediate-luminosity AGN in the COSMOS field by analyzing their environments and host galaxies.

Previous similar work employing HST survey data has found conflicting evidence.  \citet{gro05} analyzed $\sim$100 X-ray selected AGN at redshifts up to $\sim$1.3 in the GOODS fields and found no significant differences between their host structural properties and companion fractions and those measured for matched control sample galaxies.  \citet{pie07}, on the other hand, examined $\sim$60 X-ray and IR-selected AGN with $0.2 < z < 1.2$ in the Extended Groth Strip and found that AGN are marginally more likely than control galaxies to have nearby companions.  In both studies X-ray selected AGN are found to reside predominantly in host galaxies with bulge-dominated morphologies, generally in agreement with work at low and high redshift quasar host studies \citep{jah04, san04}.  Investigating larger-scale environments of 52 quasars from the DEEP2 redshift survey, \cite{coi07} showed that the quasar-galaxy cross-correlation function at $z\sim$1 closely resembles the galaxy auto-correlation function at all scales, and that the relative quasar bias traces that of blue galaxies better than red galaxies.  This might suggest that high luminosity AGN reside in blue bulges \citep{jah04, sil08} that have not yet migrated to the high-density environments typically found for massive red galaxies.

In the nearby universe, where precise techniques allow detailed studies of tens of thousands of AGN hosts (with, e.g., the Sloan Digital Sky Survey), active nuclei reside in the most massive galaxies, with structural properties similar to early-type galaxies but with relatively young stellar populations \citep{kau03}.  Examining images of $\sim$100 of the most luminous AGN within $z<0.1$, \citeauthor{kau03} found that they occupy roughly equal fractions of blue spheroids, single disk galaxies, and disturbed/interacting galaxies.  Furthermore, low-redshift close galaxy pairs that exhibit strong indications of interaction are more likely to include AGN than pairs without interaction indicators \citep{alo07}.  At the same time, black hole accretion activity is significantly larger for AGN with bright companions than otherwise \citep{alo07}, and quasars are found to have local galaxy overdensities within 100 kpc in excess of that seen in non-active galaxies \citep{ser06} and lower-luminosity AGN \citep{str08}.  Combined with earlier imaging studies showing that a minority of local Type 1 AGN are undergoing interactions \citep[e.g.][]{der98}, these studies suggest that mergers and interactions play some role in fueling AGN, but not necessarily a dominant one.

These low-redshift AGN may represent a different population than that found closer to the peak of AGN activity, $z\simeq$ 2, with different fueling mechanisms in play.  The typical AGN in a local survey like the SDSS is less luminous and possibly hosted by a galaxy in a different evolutionary state than AGN selected at higher redshift.  Intermediate- and high-redshift AGN ($z > 0.5$) are typically found in bluer, more extended, and more irregular galaxies than their low-redshift counterparts \citep[cf. ][]{jah04, san04}.  By using samples with a wider redshift range we can constrain the dominant mechanism in the history of AGN fueling, and potentially uncover its evolution over cosmic time.  Samples at $z>0.3$, however, remain limited to a few dozen objects selected through a variety of methods (see studies mentioned above).  In the present study, we extend this moderate-redshift sample and compare the effects of different selection techniques.

Using the extensive data of the Cosmic Evolution Survey (COSMOS), we determine basic properties of a sample of AGN hosts to probe their co-evolution with SMBHs.  Selection of the AGN sample and data used for the analysis is described in \S\ref{sec.sample}.  The analyses of the host galaxies are presented in \S\S\ref{sec.morph} and \ref{sec.env}, followed by a brief discussion and conclusions.  For cosmological calculations, we adopt $h=0.75$, (where $H_0=100h$ km s$^{-1}$ Mpc$^{-1}$ is the Hubble parameter), $\Omega_M=0.3$ (matter density parameter), and $\Omega_{\Lambda}=0.7$ (cosmological constant density parameter).
 
\section{Data and Sample}
\label{sec.sample}
\begin{deluxetable*}{lccc}
\tablewidth{0pt}
\tablecaption{Numbers of AGN in the samples. \label{tbl:sample}}
\tablehead{
\colhead{AGN sample} & \colhead{No. of objects \tablenotemark{a}} & \colhead{w/ ACS images} & \colhead{Successful 2-D fits} }
\startdata
AGN candidates w/ spectra & $\sim$1300 & *** & *** \\
Spec. redshifts $0.3<z<1.0$ & 459 & 391 & 314 \\
Type 1 & 48 & 36 & 19 \\
X-ray Class 2 & 83 (95) & 73 & 57 \\
X-ray Class 3 & 48 (62) & 44 & 36 \\
Radio Class 2 & 40 (120) & 32 & 30 \\
Radio Class 3 & 82 (134) & 71 & 62 \\
\enddata
\tablenotetext{a}{Numbers in parentheses include those objects in the X-ray sample with $\log(L_X)<42$, and those objects in the radio sample not classified as AGN.  Objects failing to satisfy these AGN classifications are excluded from the other columns, as well as the analyses in this work.}
\end{deluxetable*}

COSMOS \citep{sco07a}, a Hubble Space Telescope Treasury project, includes coverage of a 2 square degree field from X-ray wavelengths to UV, optical, IR, and radio.  The cornerstone data set, which we use for the bulk of our analysis, consists of 583 orbits taken with Hubble's Advanced Camera for Surveys (ACS) with the F814W filter \citep[see][for a complete description]{koe07}.  Ancillary observations include XMM-Newton X-ray imaging \citep{has07}, VLA radio maps \citep{sch07}, and VLT/VIMOS \citep{lil07} and Magellan/IMACS optical spectroscopy \citep[][Trump \etal~ 2008, in preparation]{tru07}.

Our sample selection focuses on AGN candidates in the COSMOS field with spectroscopic redshifts.  An object is identified as an AGN candidate through detection as an X-ray point source above the $\sim10^{-15}$ erg cm$^{-2}$ s$^{-1}$ flux limit in the 0.5-2 keV or 2-10 keV flux bands \citep{cappell07, bru07}, or a radio source above the 0.1 mJy flux limit at 1.4 GHz \citep{sch07}.  Optical counterparts to these candidates with $I_{AB}<24$ are followed up in the Magellan/IMACS spectroscopic survey, whose first season of observations is detailed in \citet{tru07}, including emission line identification and redshift determination for $\sim$350 objects.  We include additional sources from the second season of IMACS observations, as well as companion observations using MMT/Hectospec (Trump et al., in preparation).  Targets with successful redshift determinations are separated into 4 primary spectral classes: 1) broad emission line AGN; 2) narrow emission line objects; 3) red galaxies, with detectable continua but no emission lines; and 4) hybrids showing narrow emission lines superposed on a red galaxy continuum.

From the spectroscopically-confirmed AGN candidates, we restrict our redshift range of interest to $0.3<z<1.0$.   The upper limit arises from our limited ability to adequately analyze targets at high redshift using single-orbit ACS data, and because the ACS F814W bandpass shifts into the rest-frame UV at $z > 1$, biasing morphological characterization.  Based on our simulations described in section 3.1, typical AGN hosts at this redshift have recovered magnitude uncertainties of $\sim$0.4 magnitudes.  The lower redshift limit is a practical one, applied to limit ourselves to objects which have a large number of corresponding inactive galaxies and whose environments can be analyzed adequately within a 2 square degree field.  Additionally, some of the ancillary survey boundaries extend beyond those of the ACS observations, so $\sim$60 AGN candidates were identified for which we have no Hubble data for host analysis.  These objects are excluded from the sample.

Determining which candidates truly host an AGN is non-trivial.  Those objects with broad emission lines are the easiest to classify as AGN, but we treat them separately due to the uncertainties in analysis of their hosts (see below).  A significant fraction of the narrow-line objects may be star-forming galaxies rather than genuine AGN (Trump, private communication), and the IMACS spectra lack sufficient signal-to-noise and spectral range to make such a distinction using line-ratio diagnostic diagrams \citep[cf. ][]{bal81, vei87, kew01, kau03}.  Furthermore, many X-ray selected candidates, which are expected to be AGN \citep{mus04}, exhibit no emission lines, falling into the red galaxy spectral classification.  This might occur due to obscuration of the regions emitting optical spectral lines, misidentification of the optical counterparts, or misplacement of the slit when performing optical spectroscopy.

With these complications in mind, we present our findings separately for various sub-samples, and we use the following terms to describe them.  ``AGN candidates'' refers to the full sample of candidate AGN with spectroscopic redshifts, and includes 459 objects.  ``Class 1'' or Type 1 ``broad line'' AGN refers to those candidates with broad optical emission lines, easily distinguishable as AGN or quasars, and including 48 objects.  ``Class 2'' or ``narrow emission line'' objects refers to AGN candidates with narrow emission lines in their spectra, including 120 radio-selected objects and 95 X-ray-selected objects.  This category generally includes both Type 2 AGN as well as star-forming galaxies.  ``Class 3'' or ``non-emission line'' candidates are those whose spectra look like red galaxies with no emission lines, including 134 radio-selected and 62 X-ray selected objects.  In cases where an object has been detected both in radio and X-ray emission, we include it in the X-ray category.  These sub-samples are summarized in Table \ref{tbl:sample}.

As in previous AGN studies with similar redshift ranges \citep{pie07, sil08, bun07}, we adopt a cutoff in X-ray luminosity above which objects are likely to be AGN-dominated \citep{bau04}.  After excluding broad-line AGN, we take those candidates with $L_{2-10 keV} > 10^{42}$ ergs s$^{-1}$ as the most likely to harbor accreting black holes.  Here, $L_{2-10 keV}$ is the X-ray luminosity in the 2-10 keV band.  We estimate this luminosity for all X-ray sources by converting the observed XMM-Newton X-ray fluxes via a k-correction.  To do so, we assume an X-ray power law slope of $\Gamma = 1.9$ and perform the k-corrections using spectroscopic redshifts.  Of the 95 X-ray-selected Class 2 objects, 83 satisfy the luminosity cut given above, and we refer to them as ``X-ray Class 2'' AGN.  Of the 62 X-ray-selected class 3 objects, 48 satisfy the luminosity cut, and we refer to them as ``X-ray Class 3'' objects.  In some parts of our analysis, we combine these two samples together, and generally refer to them as X-ray AGN.  Those objects which do not satisfy the luminosity cut are excluded from the following analyses.  We note here that some X-ray point sources have more than one possible optical counterpart, with the most likely counterpart chosen.  This ambiguity applies only to three objects in our final sample, so we suspect it has little effect.

For the radio-selected AGN candidates, we use the novel technique for
separating AGN and star-forming galaxies described in \citep{smo08}.
Briefly, the technique uses a combination of morphology and rest-frame
colors of optical counterparts to the radio sources to classify them
as QSOs, stars, star-forming galaxies, AGN or high-redshift galaxies.
From the modified Stromgren photometric system \citep{ode02},
\citet{smo06} derive two principle-component combinations of
rest-frame colors which ``optimally quantify the distribution of
galaxies in the rest-frame color-color space'' \citep{smo08}.  Because
the emission line strengths correlate with galaxy spectral energy
distributions, these color parameters can mimic line diagnostic
diagrams' ability to separate emission line galaxies into AGN,
composite, and star-forming objects.  \citet{smo08} calibrate this
color classification scheme using galaxies from SDSS matched to the
NRAO VLA Sky Survey at 1.4 GHz \citep{con98}, and apply it to the
COSMOS radio sources \citep{sch07}.  The low-redshift calibration
suggests that the sample classified as AGN contatins $\sim$5\%
star-forming galaxies, $\sim$15\% composite objects, and $\sim$80\%
AGN, and these AGN comprise $\sim$90\% of the total population of
radio AGN.  When we apply the classification scheme to the
radio-selected objects in our sample, we find that 122 are identified
as AGN, including 40 Class 2 objects and 82 Class 3 objects.  We refer
to these as radio AGN throughout the paper.  Roughly 30\% of our
radio-selected AGN candidates are not classified either as AGN or
star-forming because \citet{smo08} a) use a conservative search radius
of 0.5 arcseconds when identifying optical counterparts, whereas the
spectroscopic follow-up included objects as far as 1 arcsecond from
the peak radio emission, and b) they exclude from their radio--optical
sample a fraction of objects ($15\%$) that are photometrically flagged in
the COSMOS photometric redshift catalog \citet[$\sim30\%$ of which have
available spectroscopic redshifts; see Tab.~1 in \citealt{smo08}]{cap07b}. 
The remaining objects, mostly Class 2, are classified as star-forming galaxies.

We include the broad line AGN for comparison only.  Because the redshift distribution of broad line AGN peaks at higher redshift than the Type 2 distribution, the Type 1 sample includes only $\sim$50 objects.  Furthermore, the nuclear component in broad-line AGN tends to dominate the total optical flux more than in narrow-line AGN, so some of the host-fitting techniques that we apply suffer from more serious systematic uncertainties.  These factors hamper our ability to compare with confidence the structural parameters and interaction indicators between Type 1 and Type 2 AGN hosts.  We apply the present analyses to all objects for completeness.

\begin{figure}[pht]
\plotone{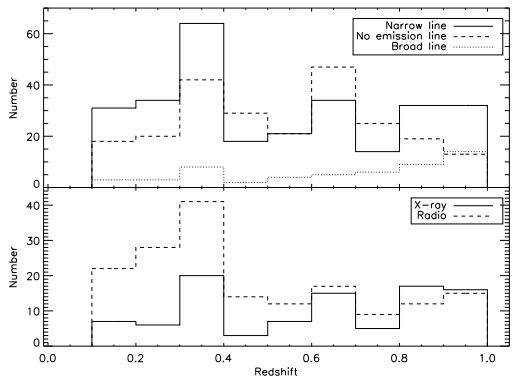}
\caption{Top panel: redshift distributions of narrow-line (solid line) and broad-line (dotted line) AGN cadidates, along with candidates with no emission lines (dashed line), from seasons 1 and 2 of IMACS observations.  Bottom panel: redshift distributions of narrow line AGN candidates selected using radio emission (dashed line) and X-ray emission (solid line).}
\label{fig.zdist}
\end{figure}
Our AGN sample thus includes $\sim$200 objects with spectroscopically confirmed redshifts in the range $0.3<z<1.0$, narrow emission line identifications, and ACS imaging.  Figure \ref{fig.zdist} shows the redshift distributions of AGN in our samples, extending to lower redshift for reference.  We suspect the redshift peaks near $z=0.3$ and $z=0.7$ are associated with large scale structures in the COSMOS field \citep{sco07b}.  Broad-line objects are plotted for comparison, and the distributions of radio-selected and X-ray-selected narrow-line objects are compared.
\begin{figure}[pht]
\plotone{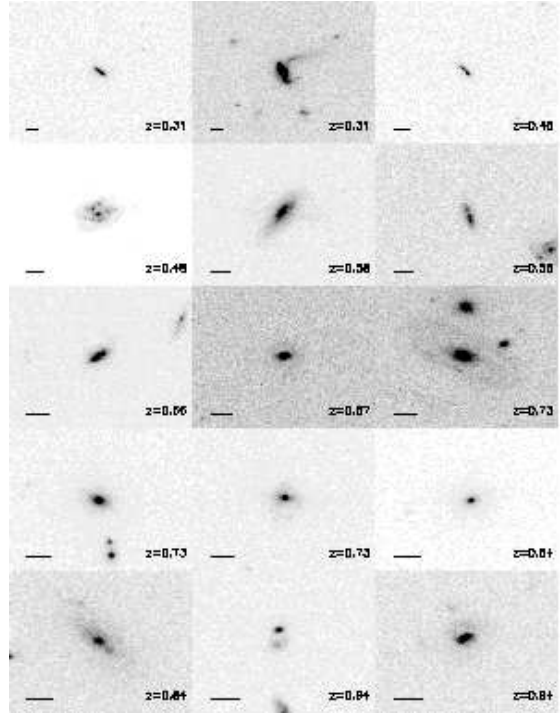}
\caption{Example Hubble/ACS F814W images of AGN with $0.3<z<1.0$.  The bar in the lower left corner of each panel is one arcsecond long ($\simeq 5h^{-1}$ physical kpc), and the pixel brightness scale is logarithmic.  We show the redshift (z) in the lower right corner.}
\label{fig.cutouts}
\end{figure}

We apply several techniques, described in the following sections, to the ACS images to analyze the AGN host galaxies.  Each image combines data from 4 sub-exposures, achieving a 5 $\sigma$ point source sensitivity of $\sim$27.2 magnitudes in $I_{AB}$ with the F814W filter \citep{koe07}.  We use the original drizzled images with 0.03 arcsecond per pixel scale, not rotated to north-up orientation, with multidrizzle parameters chosen to best preserve the original point spread function (PSF).  Figure \ref{fig.cutouts} shows example cutout images of AGN in our sample.  To complement the morphological analysis, we use the COSMOS photometric redshift catalog \citep{mob07, cap07b} as a filter to search for near neighbors to the AGN.

In order to compare the AGN sample with non-active galaxies, we also identify a control sample using the COSMOS photometric redshift catalog.  Within a survey as large and deep as COSMOS, an ideal control sample would include galaxies with the same mass and redshift distributions as the AGN.  Due to possible contamination from the nuclear point sources, we cannot use the photometric information to determine mass-to-light ratios, and thus masses, for our AGN candidates.  Therefore, we use the best-fit apparent magnitude of the AGN hosts to match the luminosity and redshift distribution of the control sample so that k-corrections can be ignored.  For each AGN in the sample, we find ten non-active galaxies with similar redshifts ($\Delta z \leq \sigma_z$, where $\sigma_z$ is the error in the photometric redshift for the control galaxies) and apparent magnitudes ($\Delta I_{AB} \leq 0.3$, using the ACS $F814W$ detections) to those of the AGN host.  We perform analyses on the control galaxies in a comparable fashion to the hosts, as described in the following sections.  For some comparisons, we find it illustrative to separate the control sample into early and late spectral types using the photometric redshift catalog $T_{phot}$ parameter, which classifies galaxies on a scale from 1.0 (red elliptical) to 6.0 (starburst) using photometric fits to the galaxy spectral energy distributions.  We divide the control sample at $T_{phot} = 2.0$, corresponding to an Sa/Sb spectral type.

\section{Morphological Analysis}
\label{sec.morph}
Using the deep, high-resolution ACS COSMOS images we attempt to determine properties of the host galaxies of our AGN sample.  The high angular resolution of Hubble's diffraction-limited imaging allows us to separate host galaxy light from that of the nucleus.  Using only the I-band images, we can constrain the magnitude, scale, radial light profile, and orientation of the host galaxy.

In this work we use 2-dimensional surface brightness fitting \citep[with GALFIT,][]{pen02} to measure AGN host properties.  To understand systematic uncertainties in the surface brightness fitting, we simulate AGN images and apply identical fitting techniques.  After decomposing the images into AGN point source and galaxy light, we measure the asymmetry and concentration of the underlying host galaxies and compare them directly with the non-active control galaxies.  We first describe our simulated AGN images, then explain the techniques we use for 2-D surface brightness fitting, and describe the results of our 2-D fits.  Finally, we discuss the asymmetry and concentration measurements.

\subsection{Simulations}
\label{sec.sims}
We performed two types of simulations to help understand systematic uncertainties in our analysis.  One suite of simulations aims to quantify our ability to reliably recover parameters in 2-D surface brightness models.  The other suite examines the effect of point-spread function (PSF) variation and mis-application on the analysis.  In this way we isolate the impacts of the two most important problems with 2-D fitting.

\subsubsection{PSF Variations}
\label{sec.psf_sim}

In performing 2-D fits to galaxy images, a PSF must be supplied to convolve with the galaxy model image.  Fits of AGN images are especially sensitive to the PSF due to the sometimes bright, nuclear point source whose light is superimposed on that of the host galaxy.  The ACS instrument's PSF ellipticity and size are known to vary both temporally and across the CCD at the level of a few percent \citep{rho07}.  Our solution, described in \S\ref{sec.fits}, includes sets of model PSF grids.  Because systematic uncertainties in the PSF can dominate the morphological classification of compact AGN hosts, we have performed a series of simulations to test how our ability to recover host properties varies with PSF.

We simulate AGN images by superimposing a real star extracted from an ACS image onto sets of simple model galaxy images with varying parameters whose ranges are similar to those of the AGN sample.  The model galaxy images are created using GALFIT, with effective radius and magnitude randomly chosen from uniform distributions in the ranges 0.15$\arcsec < r_{eff} < $ 2.5$\arcsec$ and 19 $<m<$ 24, respectively.  GALFIT convolves the specified galaxy model with chosen stellar PSF to yield the model galaxy image.  The star image is then scaled to a random magnitude with $16 < m < 25$ and added to the galaxy model image.  We create 2000 such simulated images with exponential disk profiles, and an additional 2000 with deVaucouleurs profiles.  We use four different real star images for the PSFs, with 500 simulated images per star.  One of the four stars was chosen to be near the limit of saturation, and results for the corresponding 500 simulations are obviously skewed and thus ignored in later discussion.  We refer to the simulations as ``PSF simulations,'' and we use them below to characterize the systematic effects of PSF variation on the results of 2-D surface brightness fits.  To check whether the simulations created with these four stars adequately encompass the full range of PSF variations, we also created 500 simulated AGN images using 50 different stars (10 simulated images per star), each taken from a different ACS tile and a different detector position.  As the results of fits to these images closely mimic the results obtained with the original four stars, we leave them out of the discussion below.

\subsubsection{Parameter Recovery}
\label{sec.recovery_sim}

Even if we have applied a perfect PSF, signal-to-noise limits our ability to recover fit parameters accurately, and parameter uncertainties are dominated by systematic effects.  To gauge the robustness of recovered parameters and assign appropriate uncertainties, we have performed a set of simulations where the PSF remains constant.

We simulate 2000 AGN images with exponential disk hosts, and another 2000 with deVaucouleurs hosts, with the same range of parameters as for the PSF simulations.  To better represent the background sky noise, which is the dominant noise component in our images, we randomly add cutouts from COSMOS images.  While these background images will sometimes include contaminating galaxies, the same is true for our real AGN images and the overall effect of the galaxies is minimal.  We refer to these simulations as ``Recovery Simulations,'' and by performing 2-D fits on them we characterize the uncertainties in our best-fit AGN parameters due to noise.

\subsection{2-D Surface Brightness Fitting}
\label{sec.fits}

We use GALFIT \citep{pen02} to fit models to AGN images in the sample.  For each image, we model the nuclear point source as a point spread function, and the host galaxy as a single Sersic function \citep[see][for details of the functional forms of different models in GALFIT; Sersic 1968]{pen02}.  In short, the Sersic function is a general galaxy model which encompasses a range of more specific models through the variation of an index, $n$.  The Sersic function with $n=1$ is equivalent to an exponential disk model, whereas a Sersic function with $n=4$ is equivalent to the de Vaucouleurs \cite[$r^{1/4}$; ][]{dev79} profile, which describes typical galactic bulges and early-type galaxies.  The fit results include point source position and magnitude ($m_p$), along with the host galaxy magnitude ($m_h$), effective radius ($r_h$), Sersic index ($n$), axis ratio ($b/a$), and position angle in the image.  Because some of our AGN candidates may not have a nuclear point source, we also performed fits which excluded the point source component and used just a single Sersic galaxy model.

Running GALFIT requires an initial guess of each of the best-fit parameters, an input image, a point-spread function image, and a sigma image.  Input AGN images are cut directly from the original ACS images, with a cutout image size corresponding to 35 $h^{-1}$ kpc comoving ($\sim$17'' for $z=0.3$ and 6'' for $z=1.0$; larger and smaller image sizes were attempted, with no impact on the resulting best-fit parameters).  In order to generate initial guesses in an automated way, we developed a procedure similar to that used in GALAPAGOS, described by \citet{hau07}.  First we run Source Extractor \citep{ber96} on the cutout image.  For every extracted source, we generate an elliptical mask image using the Source Extractor \verb=FLUX_RADIUS, ELONGATION,= and \verb=THETA_IMAGE= output parameters for that source.  To ensure conservative estimates of galaxy boundaries, we set the mask semi-major axis to $2\times$\verb=FLUX_RADIUS=.  Since the cutouts are centered on the AGN coordinates (which are taken as the optical counterparts of X-ray sources), we select the extracted source nearest the center of the image as the target AGN.  For each additional source in the image, we include it in the 2-D fit if and only if its mask overlaps the mask of the AGN, and otherwise we simply mask it out of the image.  All added objects are modeled as single Sersic function profiles.  Finally, we identify the brightest pixel within the AGN mask as an initial guess for the location of the point-spread function component.

For Sersic function profiles included in a fit, we estimate the initial parameters using results from Source Extractor.  The effective radius is set to $r_h = 0.162 \times$\verb=FLUX_RADIUS= based on the simulation results of \citet{hau07}.  Magnitude guesses are set to \verb=MAG_BEST=, the axis ratio determined from the \verb=ELONGATION= parameter, and the position angle computed from \verb=THETA_IMAGE=.  We constrain the Sersic index to lie between 0.5 and 8, the magnitude to stray no further than 5 magnitudes from the initial guess, and the effective radius to be less than 500 pixels (15 arcseconds).  For the AGN host Sersic component, we constrain the effective radius to be less than half the image width.  For the PSF component, we set the initial magnitude to 3 magnitudes fainter than the AGN host component (based on typical previous fits of the AGN sample).  We constrain the PSF magnitude to within 10 magnitudes of its initial value, and the position to lie within 5 pixels of its initial location.  The sky value for the image is held constant based on the sky subtraction of the original COSMOS ACS images.  We tested several methods for computing the sigma image, including conversion of the weight images output by MultiDrizzle which correspond to the ACS tiles, as well as an empirical determination of the noise based on the rms signal of regions of sky around each AGN candidate.  Differences in the choice of sigma images lead to uncertainties which are small compared to those introduced by PSF mismatch and other effects described below.

We choose the parameter constraints largely by convention, but also to ensure that they fully encompass the reasonable ranges of the parameters.  As described below, we exclude from further analysis those fits which run into constraints, since these did not find a true best-fit and the parameters are likely unphysical.  We attempted some variations on the fitting constraints; notably, we performed fits without constraints and fits where we restricted the Sersic index to $n<5$ rather than 8.  Fits without constraints fail to converge with a higher frequency than those with constraints, although this occurs mostly because our model poorly matches the real light distribution in cases of failure.  Fits with a more restricted Sersic index yield comparable results to those obtained with the original $n<8$ constraint.  A vast majority of fits with $n>5$ in the original fits ($\sim$80 objects, including 25 which run into the $n=8$ constraint) run into the constraint when we restrict $n<5$.  Placing the constraint at $n<5$ would effectively eliminate those objects from our further analysis.  However, other parameters of the fits (e.g. host magnitude) may yield reasonable and useful estimates even when a fit runs into the constraints, and these other parameters can be sensitive to the constraints chosen.  We find that the more restrictive Sersic index constraint yields host magnitude estimates systematically 0.13 magnitudes higher (dimmer) than the original constraint, with a scatter $\sim$0.5 magnitudes.  This systematic offset differs from the results of \citet{kim08} because the objects here are heavily skewed toward host galaxy-dominated images.  Furthermore, radius estimates in the restrictive Sersic index case are a median of 15 pixels (0.45 arcseconds) smaller than in the original case.  These relatively minor differences do not affect our main conclusions.
\begin{figure}[pht]
\plotone{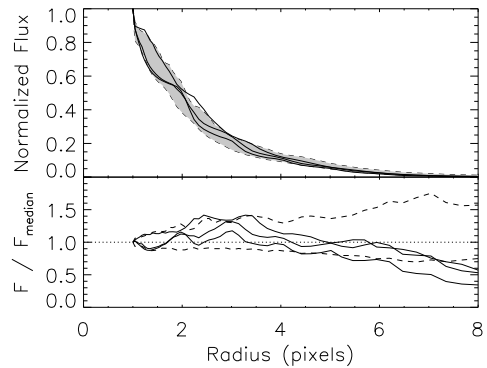}
\caption{Normalized radial point spread function profiles.  The shaded region (upper panel) and dashed lines (both panels) show the variation from the 10th percentile to the 90th percentile of profiles for 50 stars extracted from ACS images.  Thick solid lines show the profiles for three TinyTim PSF models.  In the lower panel, we divide each profile by the median profile (dotted straight line at 1.0) of the 50 real stars.}
\label{fig.psf_profiles}
\end{figure}

For the point-spread function (PSF) solution to the ACS imaging, we adopt the PSF grids described in \citet{rho07}.  These authors were motivated by the demands of detecting weak lensing signals, which require characterization of image ellipticities at the $\sim$1\% level.  Briefly, they use TinyTim software \citep{kri03} to create a PSF model at each of $\sim$4000 points in a regular grid, and develop several such grids corresponding to different focus offsets of the Hubble Space Telescope during exposure.  For each COSMOS ACS image, the best-fit focus position is obtained by simultaneously matching the shapes of model PSFs to $\sim$10 bright stars chosen from the image.  This process is found to be repeatable to an accuracy of $\sim$1 $\mu$m in focus position.  Figure \ref{fig.psf_profiles} exhibits the variation in the PSF profiles for 50 real stars selected from different ACS images at different detector positions, along with three TinyTim PSF profiles.  The TinyTim PSFs can generally encompass the variations of the real PSFs, although at large radii they systematically underestimate the flux level of real PSFs.  For an AGN at any position in an image we use the nearest model PSF from the best-fit grid.  The results of 2-D fitting of simulations described above exhibit some of the resulting systematic effects of inappropriate PSF choice.  These effects are described below in \S\ref{sec.fit_results}.

After initially fitting all the AGN candidates with GALFIT, we determined whether the results had run into the boundaries set by the parameter constraints.  Many of those objects for which this is the case have compact light profiles, so we adjusted the initial guess file such that the point source magnitude equals the host galaxy magnitude and the host radius is double the original guess, and ran GALFIT once again on those objects.  Finally, we visually inspected all of the resulting model images and residual images, subjectively assessing the quality of the fit, and in some cases attempting to remedy a failed fit.  This typically entailed masking out a nearby star or galaxy whose light was contaminating the image beyond its original mask.  We discuss fit results in the following section.

In order to place constraints on those objects for which the fits failed altogether, we used a simple point source subtraction method.  First, we fit each AGN with a single point source component and no galaxy component in GALFIT.  Then we subtracted the best-fit point source from the image.  On the residual image, we identified pixels whose flux values changed from positive on the original image to negative after subtraction (indicating over-subtraction), and set those pixels to zero flux.  Then we estimated a lower limit for the host galaxy magnitude by using aperture photometry to measure the residual flux in an aperture with a two arcsecond diameter.

We followed similar procedures with the sample of control galaxies as with the AGN candidates themselves.  We first fit the galaxies without a central point source component.  To mimic the process of fitting AGN, we then superimposed a point source and applied a fit procedure identical to the one used for the AGN candidates.  Since each AGN candidate has ten matched control galaxies, each control galaxy is matched to a particular AGN candidate.  We thus determined the brightness of the superimposed point source by using the best-fit point source magnitude from the AGN candidate fit.  Thus, the fitting performed on the control galaxies is well-matched to that performed on the AGN candidates.

\subsection{2-D Fitting Results}
\label{sec.fit_results}

\begin{figure*}[pht]
\plotone{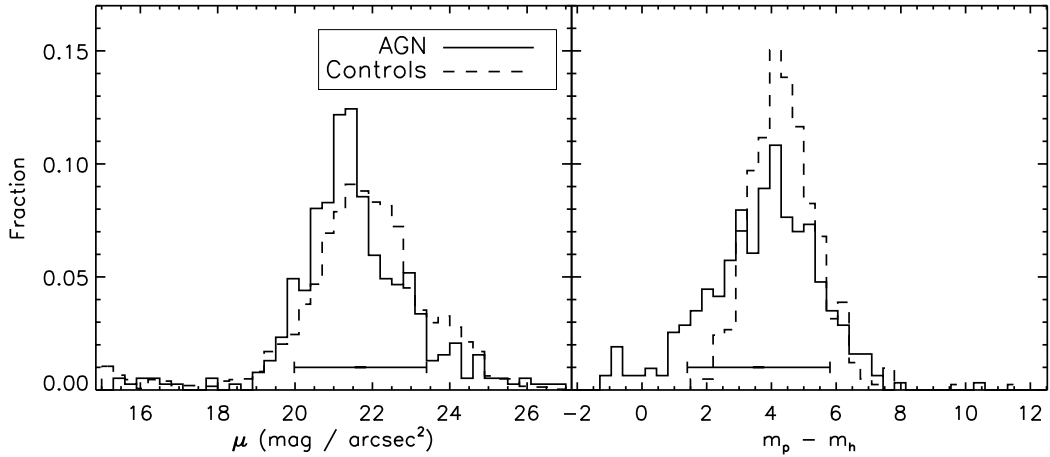}
\caption{Best-fit measured distributions of mean surface brightness (left; see text for definition) and point source to host galaxy flux ratio (right, expressed in terms of magnitudes -- objects with low values of $m_p-m_h$ have a more dominant point source than those with high values).  The horizontal bar in each plot shows the range of values encompassing the 10th to 90th percentiles of the distributions.  Distributions for control galaxies are shown as dashed lines.  The $m_p-m_h$ distribution for control galaxies comes from fits with a point source + galaxy model on inactive galaxies which should not have a real central point source.  See text for further discussion.}
\label{fig.SB}
\end{figure*}

Since the formal statistical uncertainties output by GALFIT tend to underestimate the true uncertainties, we follow \citet{hau07} and use the mean surface brightness as a proxy for image signal-to-noise to diagnose the reliability of recovered fit parameters.  The mean surface brightness is defined here as $\mu = m_h + 2.5\log(2 \pi r_h^2 b/a)$.  Figure \ref{fig.SB} shows the distributions of measured mean surface brightness for both our sample of AGN and for matched control galaxies.  The distributions have a mean near 22 magnitudes per square arcsecond, with a standard deviation of approximately 2 magnitudes per square arcsecond.  To connect to more physically meaningful galaxy characteristics, we use measurements of our large sample of control galaxies.  A typical $\sim L^*$ elliptical galaxy in the middle of our redshift range at $z=0.7$ has an effective radius $r_h\simeq 0.5$ arcseconds ($\sim 2.5 h^{-1}$ kpc) and $\mu\simeq 20.5$ magnitudes per arcsecond, while an exponential disk galaxy has $r_h \simeq 0.75$ arcseconds ($\sim 3.7 h^{-1}$ kpc) and $\mu\simeq22.1$ magnitudes per arcsecond.  In general, however, the reliability of a given best-fit parameter measurement depends in a complicated way on the other parameters in the fit.

\begin{figure}[pht]
\plotone{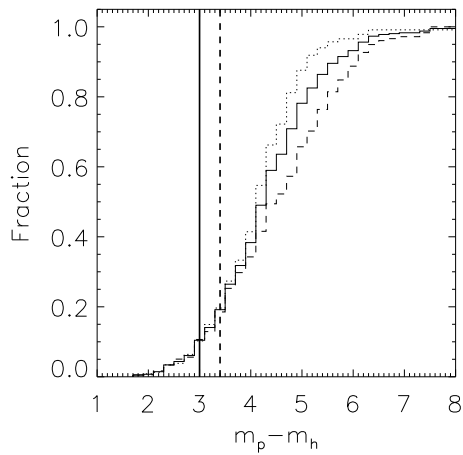}
\caption{Cumulative distribution of point source to host galaxy contrast measured for control galaxies which should lack a real point source.  Vertical lines show our chosen contrast cuts to separate real point source detections from spurious ones, in the aggressive (thick dashed, $m_p-m_h < 3.4$) and conservative (thick solid, $m_p-m_h < 3.0$) cases.  We detect point sources brighter than the conservative contrast cut in fewer than 7\% of control galaxies, so AGN candidate point sources measured to be brighter than this contrast level are unlikely to be spurious detections.  The thin dotted line shows the distribution for galaxies with late spectral types only, and the dashed early spectral types; the variation due to galaxy type is minimal.}
\label{fig.flux_ratio_cum}
\end{figure}

In particular, the inclusion of a central point source in the fit model increases the complexity of the relationships between best-fit parameters, so we consider the brightness of the galaxy component relative to the point source component as another important diagnostic.  We show this as a difference in magnitudes in the right panel of Figure \ref{fig.SB}, plotting the distribution of $m_p-m_h$ (point source magnitude minus host galaxy magnitude for the AGN candidates in the sample.  High values of $m_p-m_h$ correspond to host-dominated images, and low values correspond to point-source dominated images.  A large majority of the AGN in our sample are dominated by their host galaxies.  This begs the question: should we really include the point source component of the fit at all, or instead use just a single galaxy component?  We address the issue by fitting a sub-sample of $\sim$500 control galaxies with a galaxy + PSF model.  By including a point source component in the fit, we can determine the relative point source flux at which our fitting procedure spuriously identifies a PSF.  The distribution of spuriously recovered $m_p-m_h$ peaks near 4.5, with a broad range from $\sim$2 to 6 (dashed line in right panel of Figure \ref{fig.SB}).  We show the cumulative distribution for all control galaxies, early-type, and late-type galaxies in Figure \ref{fig.flux_ratio_cum}.  About 16\% (corresponding to the 1-sigma boundary) of these fits have $m_p-m_h<3.4$, and $\sim$7\% have $m_p-m_h<3.0$.  We establish these levels as limits to our ability to recover real point sources in AGN images.  If any given AGN fit has $m_p-m_h>3.4$ (or 3.0, more conservatively), then it is consistent with normal galaxies, lacking a real point source.  We take those AGN fits with $m_p-m_h<3.4$ (3.0) to have a real point source detection.
%

%
Formally, we can statistically determine the probability that an AGN fit with $m_p-m_h<L$ includes a real point source detection.  If we let $R$ represent a real point source detection, and $P$ represent a positive measurement that $m_p-m_h<L$, then we want to determine $p(R|P)$, the probability that an object has a real point source given a positive measurement.  From Bayes' Theorem, this is
\begin{equation}
p(R|P) = \frac{p(P|R) p(R)}{p(P)}
\label{eq.prob}
\end{equation}
We define a real detectable point source as one which yields a positive measurement, so $p(P|R)=1$.  The probability of obtaining a positive measurement depends on the false detection rate as well as the real detections.  Letting $F$ represent a lack of point source (for false detections), 
\begin{eqnarray}
p(P) &=& p(P|R) p(R) + p(P|F) p(F) \nonumber \\ 
     &=& p(P|R) p(R)+p(P|F) (1-p(R))
\label{eq.prob_positive}
\end{eqnarray}
The unknown probabilities in this equation can be estimated from our fits and the choice of $L$.  Choosing $L=3.0$, the probability of false detection is $p(P|F) \simeq 0.07$ based on fits to normal galaxies.  The probability of measuring $m_p-m_h<3.0$ can be gleaned from our fits to real AGN.  As shown in Figure \ref{fig.SB}, about 30\% of AGN fits yield a result with $m_p-m_h<3.0$.  Taking $p(P)=0.3$, we can solve Equation \ref{eq.prob_positive} for $p(R)$:
\begin{equation}
p(R) = \frac{p(P) - p(P|F)}{p(P|R) - p(P|F)} \simeq \frac{0.3 - 0.07}{1 - 0.07} \simeq 0.25.
\end{equation}
We thus use Equation \ref{eq.prob} to find $p(R|P) \simeq 0.83$, so a random AGN with $m_p-m_h<3.0$ has roughly an 83\% probability of having a real detected point source.

\begin{figure*}[pht]
\plotone{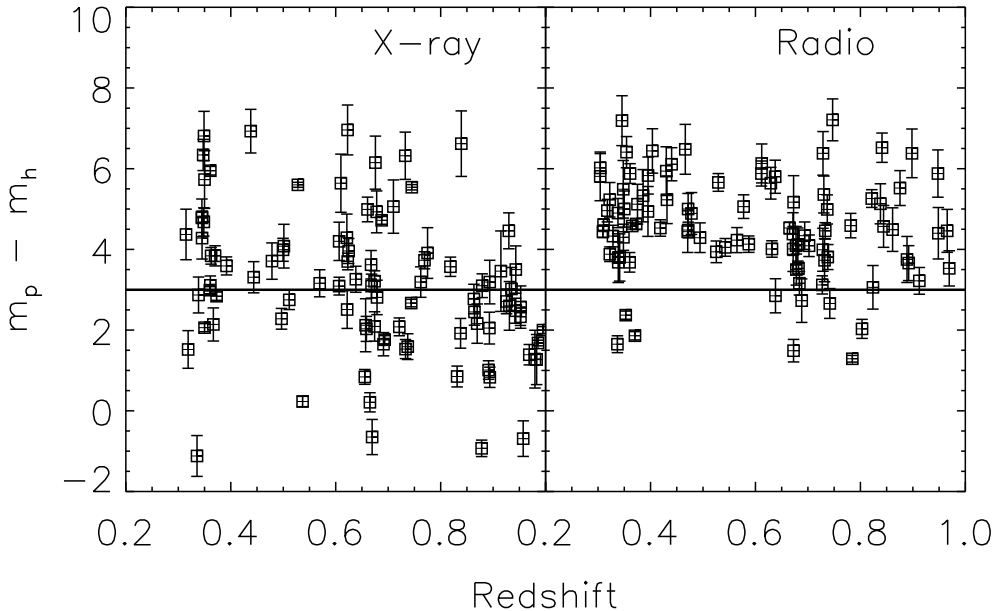}
\caption{Point source to host galaxy contrast (expressed in terms of magnitudes) as a function of redshift for X-ray selected AGN (left) and radio-selected AGN (right).  Error bars are derived from fits to simulated AGN images, as described in the text.  The thick solid line separates images with real point source detections ($m_p-m_h<3.0$) from those with likely false detections.  While X-ray AGN have relatively brighter point sources than radio objects (which are consistent with objects lacking a point source), there is no significant trend with redshift.}
\label{fig.flux_ratio_vs_z}
\end{figure*}
Figure \ref{fig.flux_ratio_vs_z} shows the host-to-point source contrast as a function of redshift for X-ray and radio AGN.  By looking at the distribution of points in the y-direction (contrast), we see that a substantial fraction ($\sim$47\%) of X-ray AGN fall below our conservative contrast cut, making them inconsistent with normal galaxies lacking a point source.  The radio AGN, however, have a distribution broadly consistent with that of normal galaxies, so they do not have detectable optical nuclear point sources.  Neither sample shows a strong trend with redshift.  This suggests that our fitting procedure, whose success (as discussed below) depends on the particular distribution of light, does not exhibit strong selection effects with redshift over the range considered here.  We show later that despite the total light output being dominated by stars, the nuclear point source can significantly impact measured morphologies.

\subsubsection{Reliability and Systematics}
\label{sec.reliability}
\begin{figure*}[pht]
\plotone{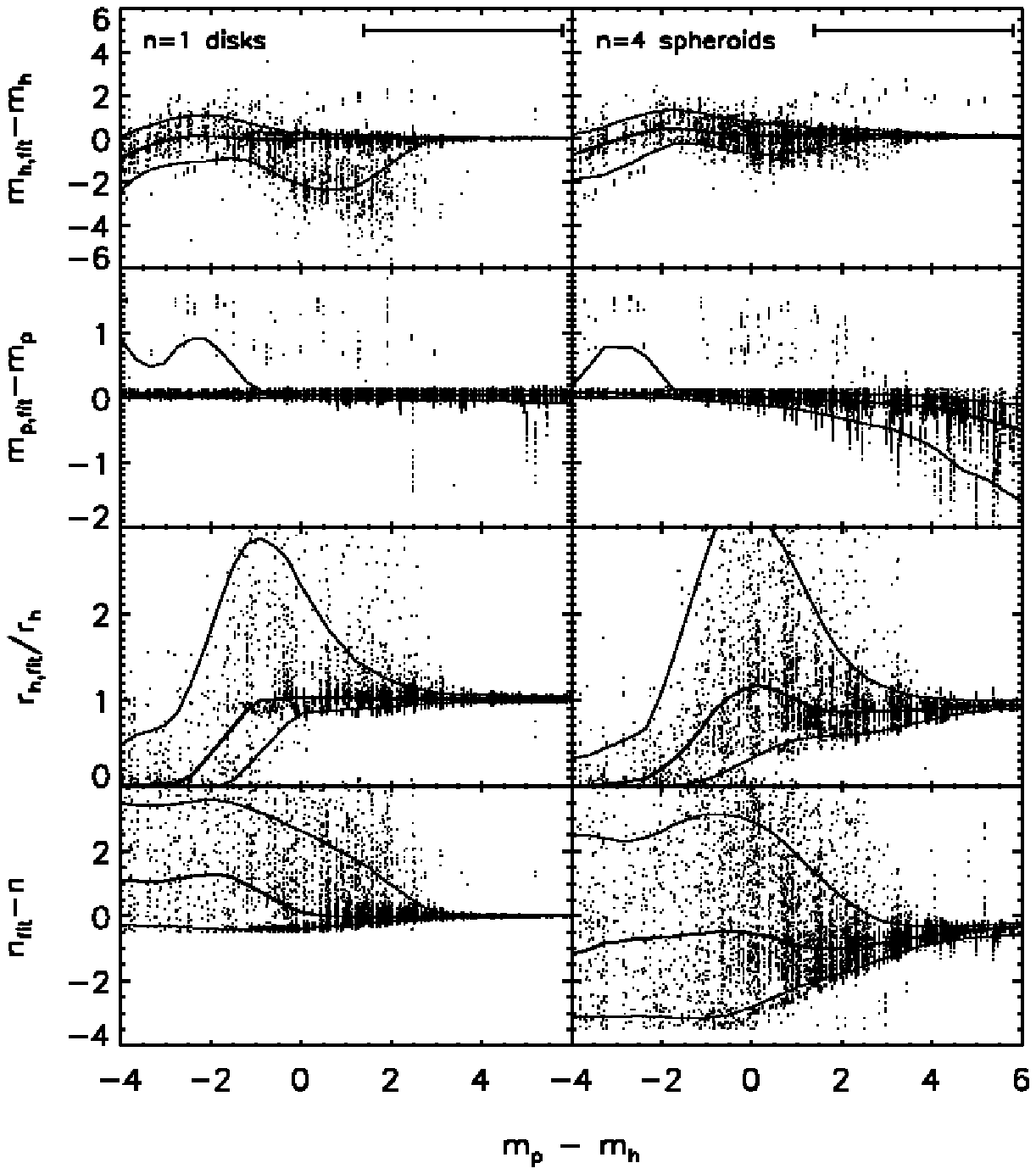}
\caption{Results of simulations gauging our ability to recover fit parameters with an imperfect PSF solution.  Each panel shows the difference between input and recovered parameter vs. input PSF magnitude minus host magnitude.  Solid lines show the 10th, 50th (median), and 90th percentiles of the recovered minus input parameter difference.  The left column shows recovered host magnitude, PSF magnitude, radius, and Sersic index for simulated AGN with exponential disks ($n=1$), and the right column shows the same for simulations with deVaucouleurs profiles ($n=4$).  The horizontal bar at upper right of each column shows the 10th to 90th percentile range of contrast measured for the real AGN in the sample.}
\label{fig.psf_sims}
\end{figure*}

We use the simulated AGN images described above to constrain systematic uncertainties in our fit results.  We fit each of the PSF simulations nine times, each time using a different model PSF taken from the grids described above.  By comparing the resulting distributions of best-fit parameters to the known original parameters, we can characterize the effect of PSF variations on our fit results.  Since we intentionally choose incorrect PSFs in this test we expect the results to be markedly worse than for our fits to real images.  Figure \ref{fig.psf_sims} shows the differences between input and best-fit parameters for the PSF simulations as a function of the difference between point source and host galaxy magnitude.  As expected, our ability to recover accurate parameters with a mis-applied PSF is generally better when the host galaxy dominates the flux of the entire system.  This does not hold for the point source magnitude for bulge-dominated systems, where the compact galaxy profile can mimic a point source.  In general, when the point source dominates the host galaxy light, the uncertainty in the fits increases with smaller galaxy radii and more compact profile.  The difficulty of recovering parameters for bright point sources is likely accentuated by the inaccuracies of the model PSF wings, which can become confused with light from the host galaxy.  Since most of the real AGN in our sample have $m_p-m_h>0$, and we have taken care to choose the best PSF, we expect these effects to be minimal.
\begin{figure*}[pht]
\plotone{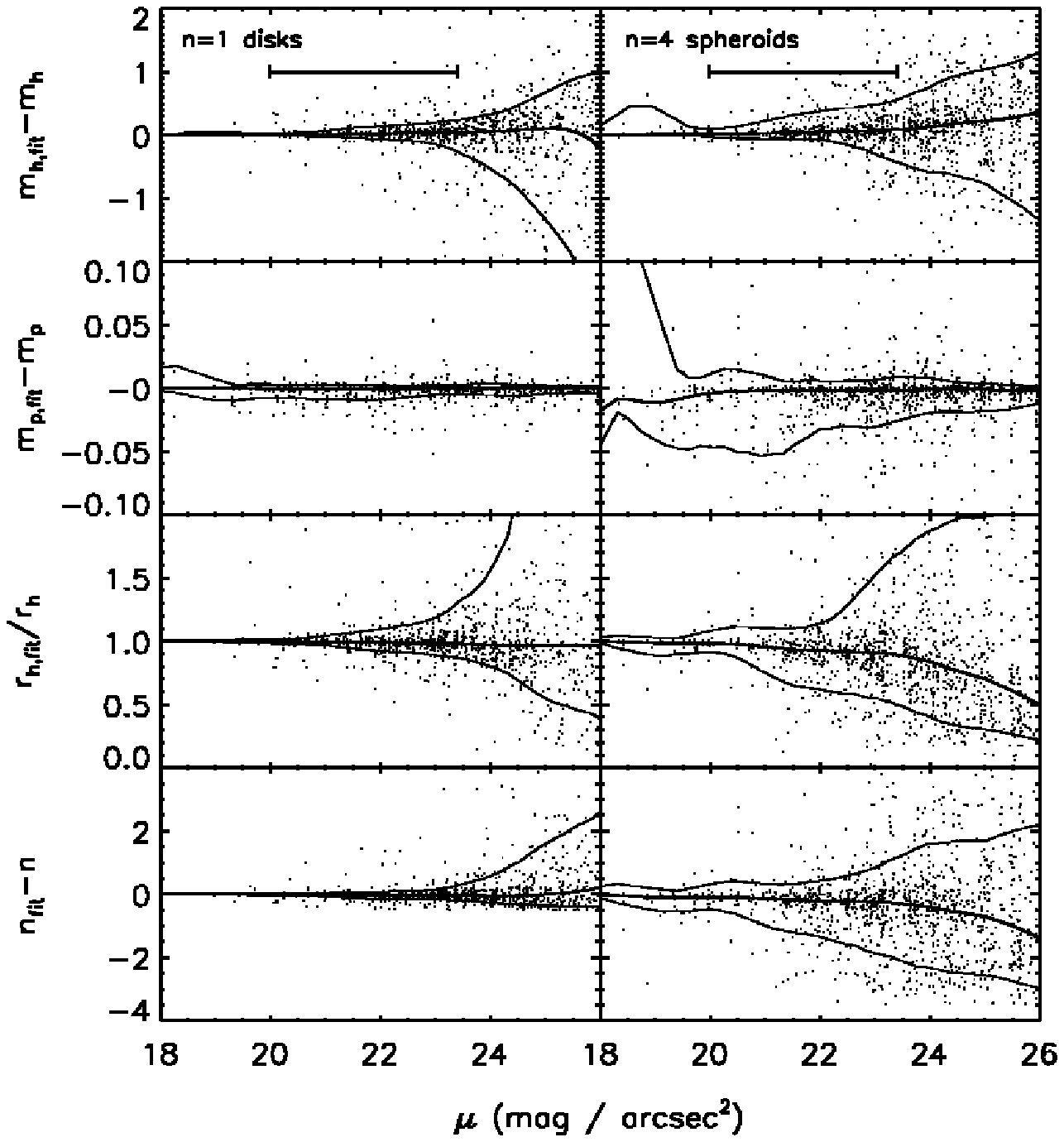}
\caption{Results of simulations gauging our ability to recover fit parameters in the presence of noise.  Each panel shows the difference between initial and recovered parameters vs. host galaxy mean surface brightness.  Solid lines trace the 10th, 50th (median), and 90th percentiles of the parameter difference.  The left column shows recovered host magnitude, PSF magnitude, host radius, and Sersic index for simulated images with exponential disk ($n=1$) profiles, and the right column shows the same for simulations with deVaucouleurs profiles ($n=4$).  The horizontal bar in top panel of each column shows the 10th to 90th percentile range of $\mu$ measured for the real AGN in the sample.}
\label{fig.recovery}
\end{figure*}

The recovery simulations help delineate the reliability with which we can recover parameters in the presence of realistic noise.  From the fitting results for these images, we determine uncertainty estimates as a function of host galaxy mean surface brightness.  We fit each image as we would a real AGN image, using the correct PSF.  Since the occurrence of galaxies is rare in the randomly selected noise images, we do not perform masking or simultaneous fitting of additional galaxy models.  We ignore fits that did not converge, and discard fit results that yield unphysical parameter values outside the boundaries set by the parameter constraints (such cases account for $\sim$15\% of the fits).  The results are shown in Figure \ref{fig.recovery}.  We clearly recover parameters better for brighter host galaxies, with substantial decreases in reliability of magnitude, radius, and Sersic index for fainter objects.  Notably, point-source magnitudes are recovered to within a tenth of a magnitude in all cases, indicating that background noise is a minor problem compared to PSF accuracy (characterized in Figure \ref{fig.psf_sims}) when measuring this parameter.

Our method for assigning realistic uncertainty estimates to the best-fit parameters of our actual AGN closely follows that of \citet{hau07}.  First, we calculate the mean surface brightness of the AGN host galaxy and take the corresponding standard deviations of the parameter distributions shown in Figure \ref{fig.recovery}.  For any one parameter, we have two uncertainty estimates, $\sigma_{n=1}$ and $\sigma_{n=4}$, corresponding to the exponential disk and de Vaucouleurs simulations.  We linearly interpolate between these two values to match the measured value of $n$, using the limiting values instead of extrapolating for $n>4$ and $n<1$.  Because our objects are mostly host-dominated, we do not attempt to determine how the uncertainty estimates vary with the point source brightness.  The additional scatter due to the range in relative point source brightness is already folded into the scatter in Figure \ref{fig.recovery}.

The simulation fit results do help us to identify systematic effects of a bright nuclear point source.  Qualitatively, a bright central PSF causes best-fit values of $r_h$ to be systematically low, $n$ to be systematically low, $m_h$ to have additional scatter, and $m_p$ to have less scatter.  These effects are pronounced primarily for objects with $m_p-m_h<0$, which excludes the bulk of our sample.

\subsubsection{Morphologies from 2-D Fits}
\label{sec.2dfit_morphs}
\begin{figure*}[pht]
\plotone{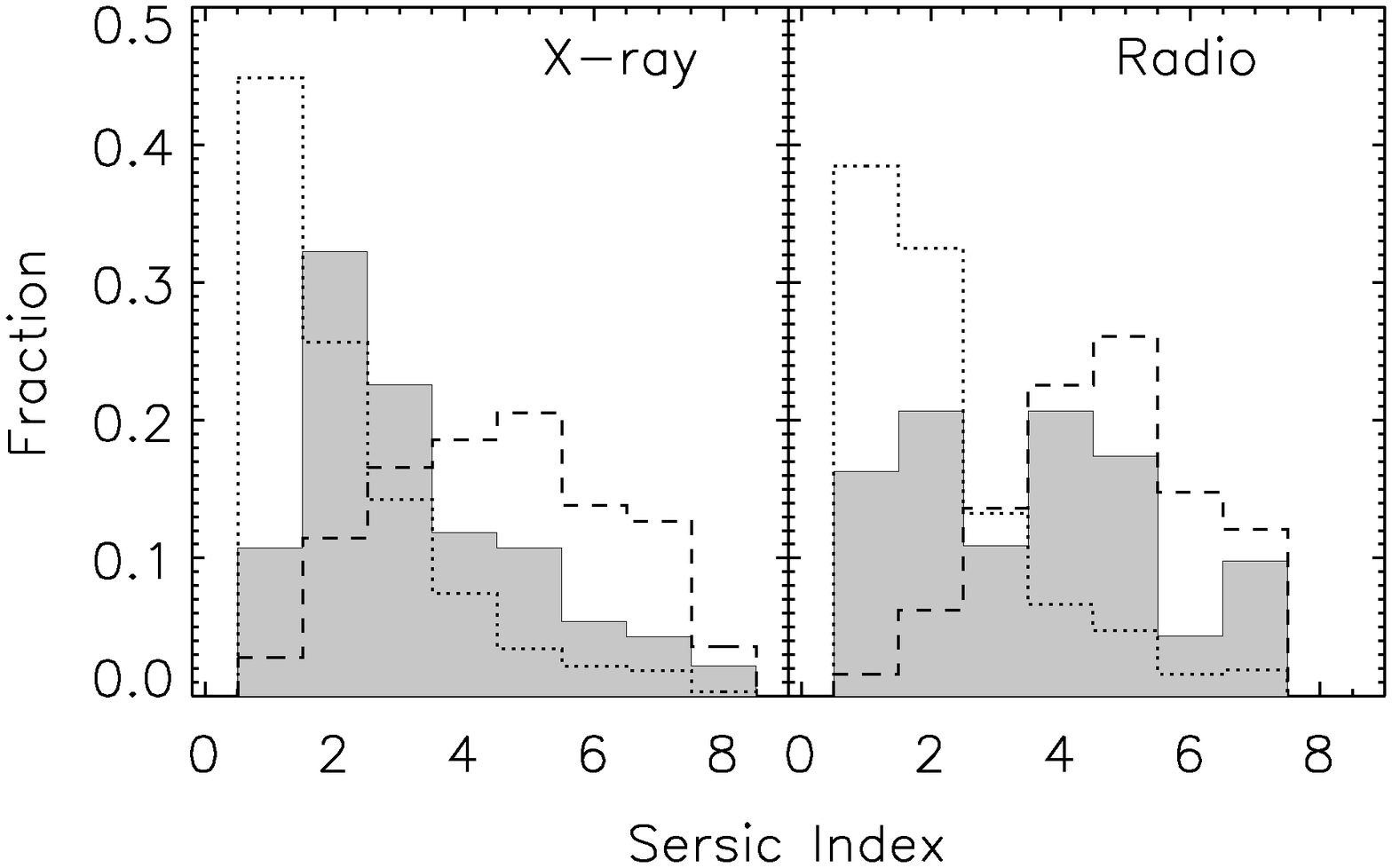}
\caption{Distributions of measured Sersic index for X-ray (left panel) and radio (right panel) AGN, along with matched control galaxies.  Control galaxies are separated into late (dotted line) and early (dashed line) spectral types based on photometric redshift fits.  X-ray AGN (shaded, left panel) include Class 2 and 3 objects (with and without narrow lines), and exhibit a morphology distribution intermediate between disk- and bulge-dominated.  The distribution for radio AGN (shaded, right panel) appears bimodal, with a significant late-type morphology component.  However, contamination of the radio sample by star-forming and hybrid galaxies, combined with the uncertainties in our fits, prevents us from distinguishing the radio AGN morphology distribution from that of early-type galaxies.}
\label{fig.sersic}
\end{figure*}

Table \ref{tab.fits} gives the results of 394 2-D fits for the AGN candidates in our sample, including all objects with ACS images in all classes.  Initially, 174 fits ran into the fit constraints and we tried again with new parameter guesses (see discussion of our fitting methods above).  Of the re-fits, 74 were successful.  After visually assessing the quality of the fits, we attempted an additional 74 fits with initial parameter guesses adjusted manually, of which 26 yielded acceptable results.  Finally, we checked whether 2D fits without a point source component were successful when the fit with a point source failed -- eight of those objects whose best fit parameters had run into constraints with a PSF component did not run into constraints without.  The values in the table are the best best-fit parameters.  We assign a flag to each fit, with the following values: 0) successful automated fit; 1) fit ran into parameter constraints; 2) automated fits ran into parameter constraints, but manual fit did not; 3) fit does not include a point source component (because inclusion of a point source led to a poor fit); 4) visually assessed as a poor fit.  We consider those objects with flag values of 0, 2, and 3 as successful.  Table \ref{tbl:sample} shows the numbers of AGN with successful fits in each of our defined classes.  In the following section we discuss asymmetry and concentration of the AGN, which can be measured even for those objects whose 2-D fits have failed.  Such objects have higher values of asymmetry (by $\sim$40\%) but similar values of concentration compared to those with successful fits.  It is not surprising that those objects with the greatest degree of asymmetry are the most difficult to fit with simple galaxy models.

Figure \ref{fig.sersic} shows distributions of the best-fit Sersic index for X-ray and radio objects and their control samples.  Class 2 objects have a distribution of Sersic index which is statistically indistinguishable from Class 3 objects, so we combine them here.  For all AGN candidates with measured $m_p-m_h>3.0$ and successful no-PSF fits, we replace the Sersic index measured when including a point-source component with that measured without including a point source.  This replacement, which accounts for those AGN without a strong nuclear point source, ultimately affects the overall distribution of Sersic index minimally.  Control galaxies are separated into late and early spectral types using the COSMOS photometric redshift catalog \citep{mob07} $T_{phot}$ parameter, and we find a good corresponding separation of morphologies into disk- and bulge-dominated.  Early-type control galaxies are clustered around $n=4$, although with significant scatter, and late-type galaxies around $n=1$.  With this division, approximately 60\% of all the control galaxies have late-type morphologies and spectral types, and the remaining 40\% have early-type morphologies and spectral types.

X-ray AGN have a Sersic index distribution intermediate between the late- and early-type control galaxies, including a broad range of morphologies.  When the control galaxies are not separated by spectral type, a KS test rejects the hypothesis that the X-ray AGN distribution is consistent with that of the controls at the 97\% level.   This result, which conflicts with some previous findings \citep{gro05, pie07}, deserves a fair amount of scrutiny.  In particular, our simulations show (cf. Figure \ref{fig.recovery}) that a substantial number of bulge-dominated AGN will have a recovered Sersic index with $n<2.5$, the typically used cutoff between disk- and bulge-dominated.  Quantitatively, $\lesssim$30\% of recovered Sersic indices will incorrectly have $n<2.5$.  On the other hand, 43\% (40/93) of the AGN in our measured distribution have $n<2.5$, indicating a significant disk-dominated population.  Furthermore, systematically low values of measured Sersic index are more prevalent for point-source dominated images included in our simulations, but most of our X-ray AGN are actually host-dominated.  Another possible effect emerges from the fact that 20\% of our X-ray AGN did not yield successful 2-D fits.  Perhaps the objects with failed fits are exactly the ones which would fill in the bulge-dominated portion of the distribution.  Our simulations show, however, that disk-dominated systems are more likely to fail than bulge-dominated ones for all values of the mean surface brightness.  Finally, we show in the following section that measurements of the concentration of these AGN reinforces the trend.

Like the X-ray AGN, radio AGN have a Sersic index distribution spanning a range of morphologies, with an apparent bimodality.  In this case, 37\% (34/92) objects have $n<2.5$.  We emphasize, however, that the radio AGN classification scheme \citep{smo08} admits $\sim$20\% contamination from star-forming and hybrid galaxies.  This effect might help explain the apparent bimodality, since the radio objects classified as star-forming galaxies have disk-dominated morphologies.  With this consideration, we cannot rule out that the distribution of morphologies for radio AGN is consistent with that for early-type control galaxies.  For none of the samples do we detect any evolution of morphology with redshift.

Because only 18/34 broad-line AGN images were successfully fit (due to difficulties fitting point source-dominated objects), we cannot make strong claims regarding their host galaxies' morphologies.  The measured Sersic index distribution favors disk-dominated morphologies, but our simulations indicate that best-fit parameters are highly uncertain for such point-source dominated objects, especially with an imperfect PSF.  We used the PSF-only fits to determine which objects have resolved host galaxies in the ACS images. To calibrate this method, we apply the PSF-subtraction technique on $\sim$60 stars selected from different COSMOS ACS tiles.  When we measure the residual flux through an aperture after subtraction, we find that residual flux is $>1$\% of the total (pre-subtracted) flux for 16\% of the stars.  The residual flux is $>5$\% of the total flux for only 3\% of the stars.  Thus, we use a 5\% flux cutoff (i.e. if the flux in residuals is above 5\%, then we claim a host galaxy detection, and if the flux in residuals is less than 5\%, then we do not claim detection) and expect to have false detections $\sim$3\% of the time.  This sort of subtraction technique is conservative in the sense that it almost always over-subtracts the PSF, yielding a lower limit on the residual flux attributed to the host galaxy.  Using this 5\% tolerance, all of the broad line AGN in our sample have resolved host galaxies, though some at a marginal level.  Table \ref{tab.type1} shows our measured limits on PSF magnitude and host magnitude for these objects, compared with the best-fit quantities taken from Table \ref{tab.fits}.

\subsection{Asymmetry and Concentration}
\label{sec.asym}

We use the asymmetry parameter, $A$, and concentration, $C$, to further quantify AGN host morphologies.  These model-independent indices (along with clumpiness) have been used as ``fundamental'' properties to classify galaxy structure \citep{abr94, abr96, con00}.

Structures with low spatial frequencies (large scales) dominate the asymmetry index, with $<30\%$ of a galaxy's asymmetry arising due to star formation \citep{con03}.  Therefore large asymmetries serve as good indicators of recent merger activity, with 50\% of nearby ULIRGS (expected to be merging systems) showing a 3-sigma deviation from the asymmetry trend with colors for normal galaxies \citep{con03}.  A conservative minimum asymmetry for merging systems is $A=0.35$, but we do not apply this limit here because we are interested only in a difference in asymmetry between active and non-active galaxies.

In this study, we compare asymmetries measured for AGN hosts to those measured for control galaxies to determine whether AGN activity is more likely to be associated with mergers and interactions.  \citet{gro05} use similar logic in applying asymmetry measurements.  Because only one filter of ACS data is available for most objects in our sample, we probe different rest wavelengths as a function of redshift.  \citet{cap07}, using COSMOS ACS images in both the F814W band as well as the F475W band (which was used to image $\sim$81 square arcminutes), find that asymmetries are systematically different when the F475W band samples rest frame UV and the F814W band samples rest-frame optical light.  Measured values are consistent, however, when both bands sample optical light or both sample UV light.  The authors illustrate that the shift in measured asymmetry values for the F814W filter occurs near $z\simeq$1, where rest-frame UV begins to dominate.  Similarly, \citet{san04}, using Sersic index to classify quasar host galaxy morphologies at $z\simeq$1, found that most objects' optical and UV classifications were the same.  We therefore expect only small systematic effects due to band shifting in the present study.

We follow the method of defining and measuring asymmetry given in \citet{con00}.  Starting with an image cutout with flux distribution $S$, we rotate the image by 180 degrees to get a new image, $S_{180}$, and define asymmetry as $A = \min (\sum \left| S-S_{180} \right| / \sum \left|S\right|) - A_0$.  The sum is over all pixels, and we take the minimum asymmetry value from a grid of central pixels near the center coordinate of the image.  $A_0$ is the asymmetry of the background, estimated by taking a median of 25 images surrounding the primary target.  The images used in constructing the background are taken from the same ACS tile as the primary, and each has the same size as the primary cutout image.  For primary targets near the edge of a tile, we shift the grid of 25 images so that all images fall within the tile's field of view.

To measure galaxy asymmetry meaningfully for a range of redshifts, we must carefully choose the size of the image cutout which we rotate and subtract.  A simple choice would be a constant physical radius, which translates directly to an angular size given a chosen cosmology.  Since galaxies come in many sizes, perhaps a better choice is to use a Petrosian radius \citep{pet76}, or a multiple thereof, as in \citet{con00}.  The Petrosian $\eta$-function, $\eta(r)$, is defined as the ratio of surface brightness at radius $r$ (from the galaxy centroid) to the average surface brightness within $r$.  We then denote the Petrosian radius as $r_{\eta_0}$, the radius at which $\eta(r) = \eta_0$.  We choose $\eta_0=0.2$ and measure asymmetries for image cutouts with this half-width.  Because the Petrosian radius can give unphysical values for unusual light distributions or for images with multiple objects, we set a minimum cutout size of 1 arcsecond and a maximum cutout size corresponding to a physical radius of 15 $h^{-1}$ kpc ($\sim 3$ arcseconds at $z=0.7$).  These restrictions prevent unrealistically small (e.g., less than the PSF full width at half max) or large choices of cutout radius.
\begin{figure*}[pht]
\centering
\plotone{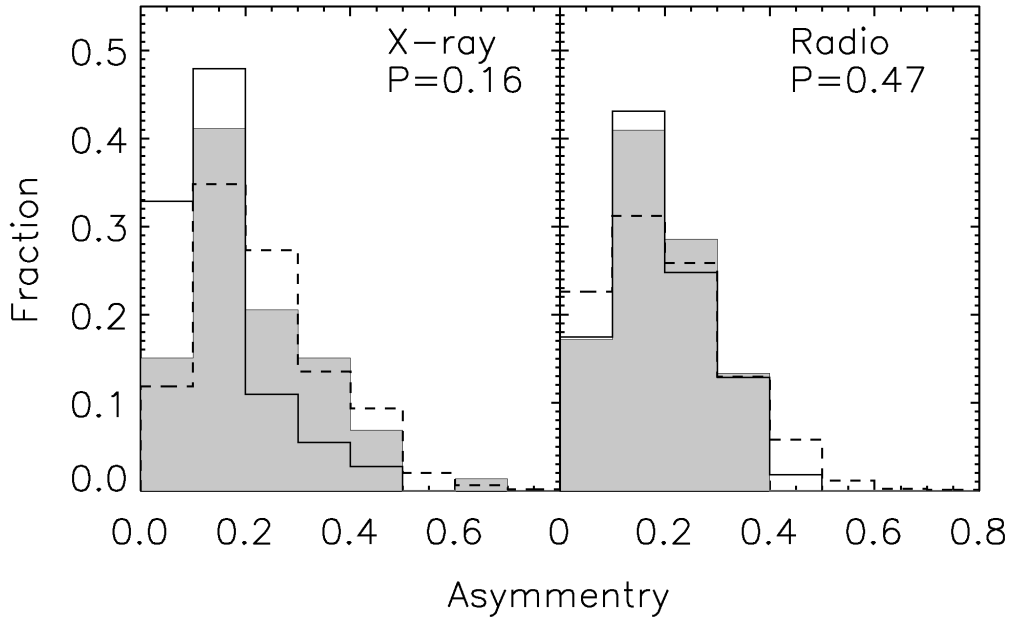}
\caption{Asymmetry distributions for Class 2 X-ray and radio AGN samples (shaded) with matched control samples (dashed).  Results for the AGN before subtraction of the best-fit point source component are shown as a thin solid line.  K-S test probabilities that the AGN asymmetry populations are drawn from the same distributions as their controls are shown at upper right in each panel.}
\label{fig.asym}
\end{figure*}

We measure asymmetries for both the AGN host galaxies and the sample of control galaxies.  Because the highly symmetric central point source of an AGN biases the asymmetry toward low values, we measure asymmetry for images with the point source component subtracted.  We subtract the best-fit model nuclear point source from each AGN image.  For objects without successful fits, we use residual images from our PSF-only-fit subtraction.  The resulting $A$ distributions for X-ray and radio AGN are shown in Figure \ref{fig.asym}, including measurements both before and after subtraction of the point source.  Point source subtraction clearly biases the results for the X-ray objects toward lower asymmetry, but very little for the radio objects.  We perform a two-sided K-S test to determine whether the AGN and control sample populations are consistent with the same underlying distribution.  We find no evidence that AGN have different asymmetry distributions from non-active galaxies, with K-S test probabilities of 16\% and 47\% (where a typical tolerance of 5\% is used to claim the distributions differ).  The asymmetries for AGN are generally consistent with those of non-active galaxies.  We find no correlation between X-ray luminosity and asymmetry.
\begin{figure}[pht]
\centering
\plotone{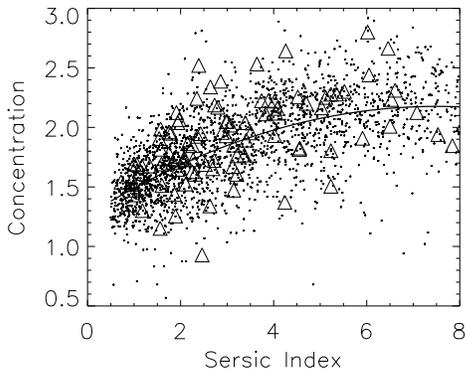}
\caption{Measured concentration, $C$, and best-fit Sersic index, $n$, for control galaxies (small dots) and X-ray AGN (triangles).  The solid line shows a best-fit parabola to the control galaxy data only.  See text for details.}
\label{fig.conc_corr}
\end{figure}

The concentration parameter serves as an alternative to the Sersic index to determine whether a galaxy is dominated by a highly concentrated central bulge component.  Here we define the concentration as $C=5\log(r_>/r_<)$, with $r_>=0.9 r_{\eta_0}$ and $r_<=0.5 r_{\eta_0}$.  Figure \ref{fig.conc_corr} shows the relationship between concentration and Sersic index for those control galaxies with successful 2D fits.  Despite the substantial scatter, the overall correlation is clear.  Because the relationship appears to flatten out for $n>4$, we fit a parabola to the control galaxy data, with the best-fit equation yielding $C = 1.262 + 0.244n - 0.0163n^2$.  For comparison, we plot the X-ray AGN in our sample, showing that the trend is comparable.  A parabolic fit to just the X-ray AGN data yields fit parameters consistent with those given above.  With our definition of $C$, our best-fit parameters show that a delineation between late- and early-type of $n=2.5$ corresponds to $C=1.8$.
\begin{figure*}[pht]
\centering
\plotone{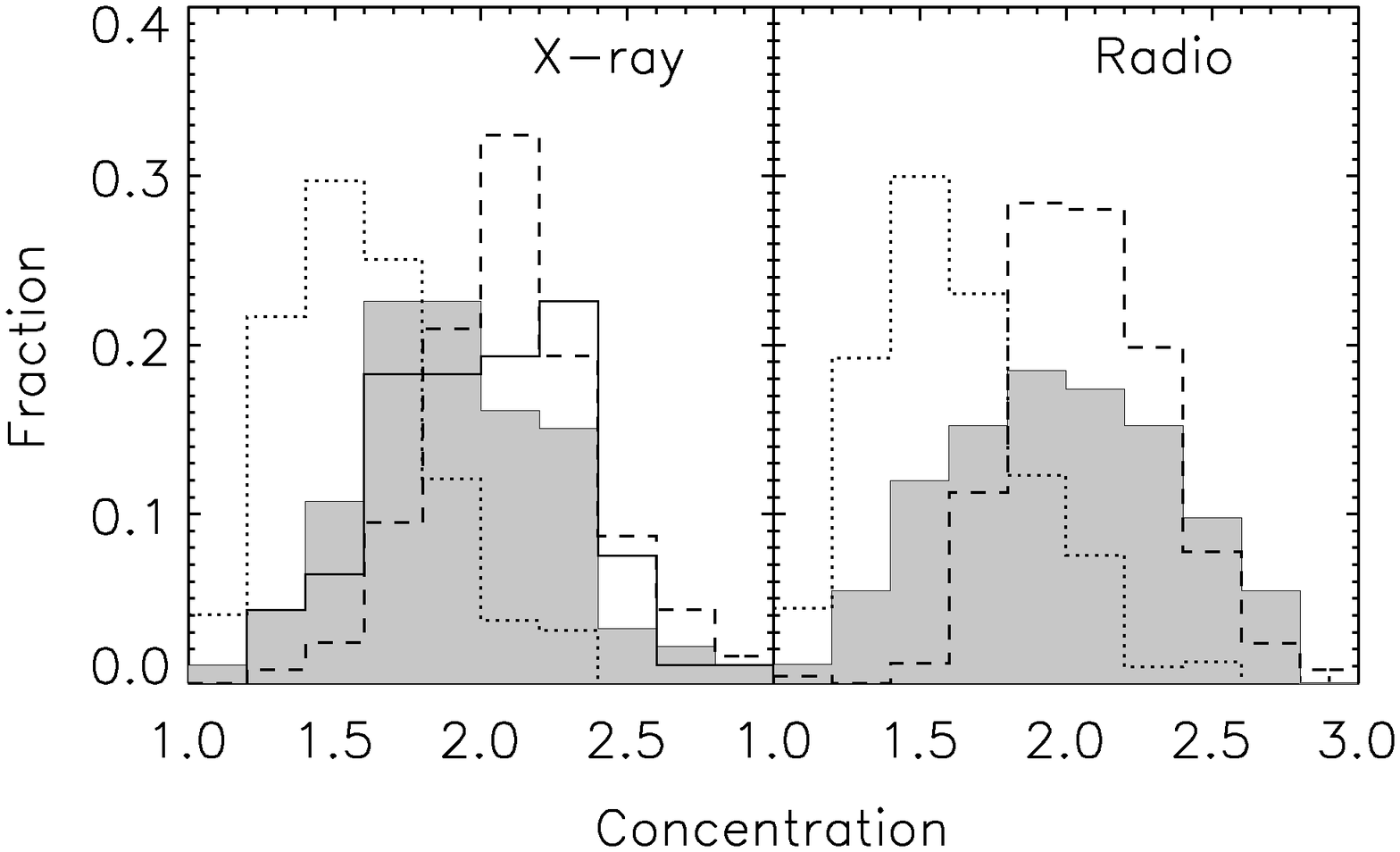}
\caption{Distributions of measured concentration, $C$, for X-ray and Radio AGN, along with control samples.  Control galaxies are separated into late (dotted line) and early (dashed line) types.  For comparison, we show the concentration distribution of X-ray AGN before point-source subtraction (thin solid).  Since the radio AGN are all host dominated, we do not show results for measurements before point-source subtraction in the right panel.  These results mimic those of our 2-D fits, as shown in Figure \ref{fig.sersic}.}
\label{fig.conc_plot}
\end{figure*}

Figure \ref{fig.conc_plot} shows the concentration distributions for X-ray and radio AGN.  As in our Sersic index analysis, we separate control galaxies into early and late spectral types.  Both X-ray and radio samples include objects both with and without emission lines combined into one.  We also show the distribution of $C$ values before (thin solid) and after (shaded) point-source subtraction for the X-ray AGN.  The presence of the point source significantly biases concentration measurements to high values for these objects.  X-ray AGN have intermediate values of $C$ between those of late and early type control galaxies.  Radio AGN also include objects with values of $C$ lower than that measured for early-type galaxies, but again we caution that the radio AGN sample includes substantial contamination. These results support those of the Sersic index distributions discussed above.

\section{Companion Galaxies}
\label{sec.env}

\begin{figure}[pht]
\centering
\plotone{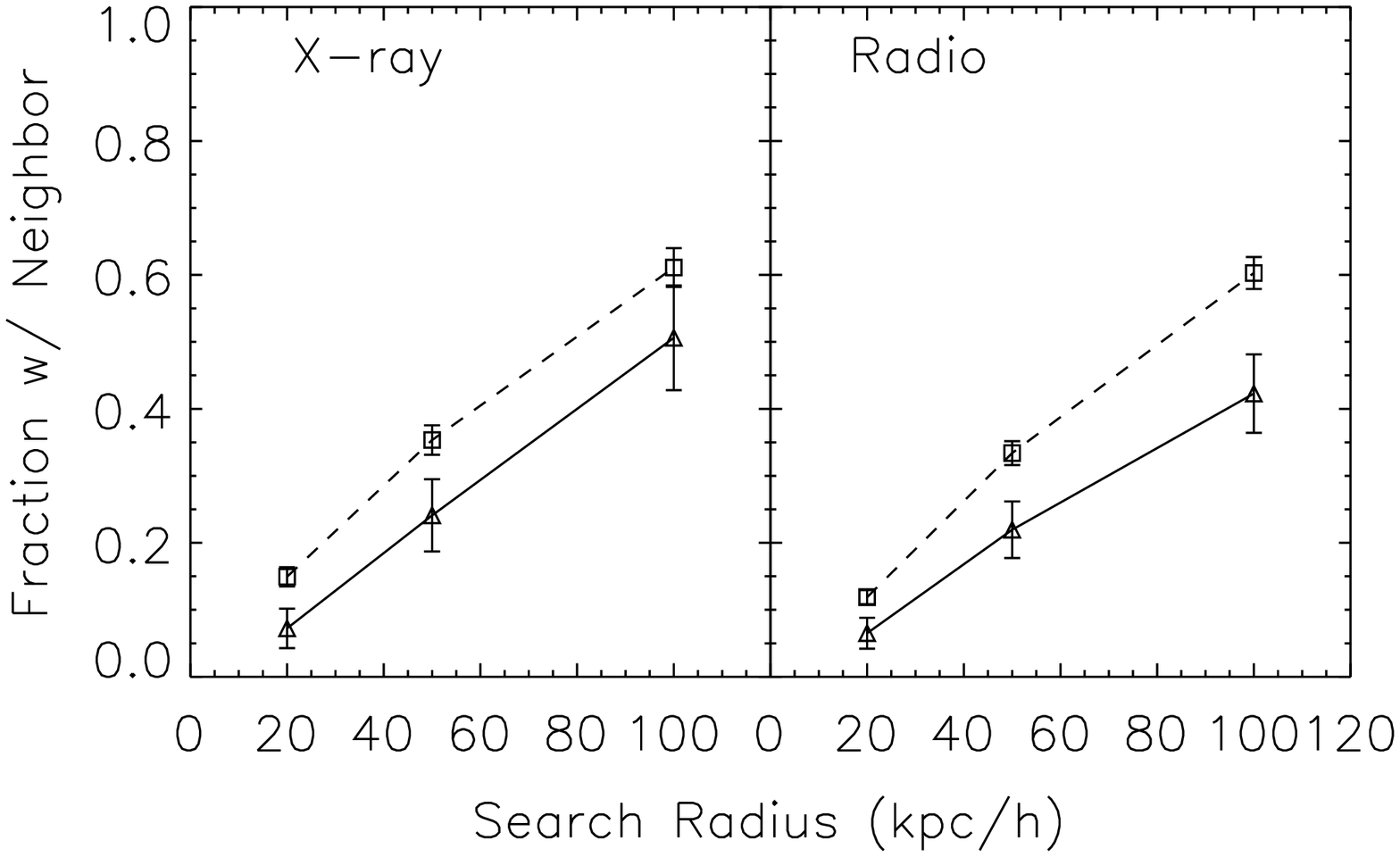}
\caption{Fraction of AGN (triangles) with at least one neighbor within a physical separation of $R_p^{max}$, as a function $R_p^{max}$.  Control galaxy neighbor fractions are shown as boxes connected by dashed lines.}
\label{fig.neighbors}
\end{figure}

Kinematically associated neighboring galaxies provide evidence for ongoing galaxy interactions.  Without detailed spectral information, we are limited to counting neighbors that are within long cylinders seen in projection, but with photometric redshift estimates and sufficient statistics we should be able to discern significant differences among samples.  \citet{pat00} prescribe a detailed method for counting kinematic neighbors with limited redshift information in flux-limited surveys.  However, our primary interest is not the absolute number of companions per galaxy, but rather the fraction of AGN with a potential physical companion relative to that number for normal galaxies.  We therefore circumvent the need for the weighting schemes described by \citet{pat00} by carefully choosing our normal galaxy sample and neighbor criteria.

We define the maximum projected physical separation of close pairs as $R_p^{max}$, and use $R_p^{max} = 20, 50, 100 h^{-1}$ kpc in three separate trials \citep[see][]{pat00, pat02}.  Due to the aperture size of the photometric catalog, neighbors within about 1.5 arcseconds ($\sim 7.5 h^{-1}$ kpc at $z=0.7$) will not be distinguished from the primary galaxy, but we expect the asymmetry measure to reflect such close companions.  We exclude a candidate neighbor galaxy if its photometric redshift is greater than 1$\sigma_z$ from that of the primary galaxy, where $\sigma_z\simeq 0.04(1+z)$ is the uncertainty in the calibrated photometric redshifts \citep{mob07}.  We must then impose a minimum luminosity for a candidate neighbor to qualify as countable \citep{pat00}.  Using the best-fit apparent magnitude of the AGN host and control galaxies, and the COSMOS photometry for the secondary galaxies, we exclude neighbor galaxies more than $dm$ apparent magnitudes ($I_{AB}$) fainter than the primary.  A choice of $dm=2$ should restrict our counting to include only neighbors which could undergo a relatively major merger with the primary.  Other choices of $dm$ yield comparable results.  An alternative to this apparent magnitude limit is an absolute magnitude limit tied to $M_V^*$, the break in the V-band luminosity function.  We tried counting neighbors with this kind of cut as well, with similar results.

Figure \ref{fig.neighbors} shows the fraction of X-ray and radio AGN with a neighboring galaxy within three different search radii.  Note that the higher search radii are inclusive of the lower ones, so the points plotted are not independent.  Error bars are estimated for the Poisson case, with $\sigma_{N}=N^{1/2}$.  We do not find significant differences between the AGN and their control galaxies' neighbor fractions.  We note that the AGN neighbor fraction appears consistently lower than that for control galaxies in the figure, but changes in the redshift tolerance and magnitude cut can reverse this effect.  No significant trends with morphology or luminosity can be discerned with the sample size used here.  We conclude that AGN are no more likely than non-active galaxies to have a near neighbor.  With spectroscopic redshifts from the COSMOS VLT survey \citep{lil07}, future work will more accurately identify kinematic neighbors.

\section{Discussion}
\label{sec.discussion}
\begin{figure}[pht]
\centering
\plotone{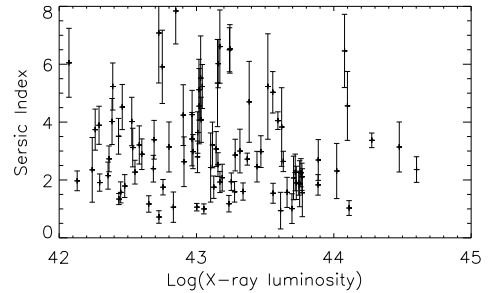}
\caption{Measured host galaxy Sersic index as a function of X-ray luminosity for X-ray AGN.  No significant trend is discernible.}
\label{fig.sersic_vs_xlum}
\end{figure}
\begin{figure*}[pht]
\centering
\plotone{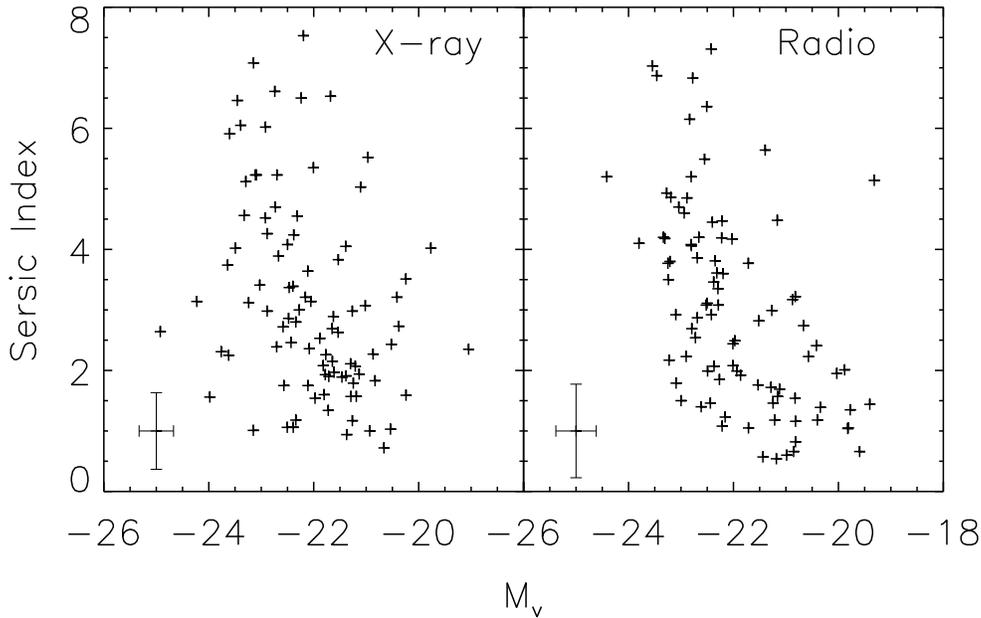}
\caption{Measured Sersic index as a function of host galaxy absolute V magnitude for X-ray and radio AGN.  Median error bars for the measured quantities are shown in the lower corner of each panel.  Magnitude uncertainties are derived solely from uncertainties in the 2D fits.  Radio objects show a slight trend for brighter host galaxies to have more bulge-dominated host galaxies.  No significant trend is found for the X-ray objects.}
\label{fig.sersic_vs_hostlum}
\end{figure*}

Recent evidence suggests that galaxies hosting AGN represent a transitional population, passing from the blue cloud to the red sequence in galaxy color-magnitude space at redshifts $z\lesssim1$ \citep{jah04, san04, sil08, bun07}.  AGN host galaxies at both low and high redshift are found to be bluer than quiescent elliptical galaxies \citep{kau03, jah04a, jah04, san04}, indicating recent or ongoing star formation.  Contrary to the findings of some previous authors studying morphologies of moderate-redshift X-ray selected AGN \citep{gro05, pie07} and quasars at low redshifts \citep{dun03, mm01}, our results indicate that X-ray AGN hosts may be undergoing a morphological transition concurrent with a transition from blue to red colors.  \citet{pen06c} came to a similar conclusion when studying the host galaxies of gravitationally lensed quasars -- 30-50\% of quasar hosts in their $z>1$ sample have disk-dominated morphologies.  Further, our results qualitatively agree with those of \citet{kau03}, whose AGN sample had a distribution of concentration index intermediate between that of early- and late-type galaxies.  In the emerging merger picture developed through simulations \citep{spr05, hop06}, a morphological transition might make sense as we witness evolutionary phases of a merger-triggered AGN.  As the merger of two gas-rich disk galaxies proceeds, accretion onto the central black hole would power the AGN as the host galaxies disrupt and form into a bulge-dominated elliptical.  \citet{has08} provides another possible scenario where bulge-dominated galaxies accrete new gas from their cosmic surroundings, leading to the build-up of a disk component and subsequent feeding of the previously dormant black hole.

The discrepancies between AGN host morphology studies likely arise from a combination of sample selection effects and biases in analysis techniques.  We have already addressed one key bias, that arising due to the presence of a nuclear point source in many optical images of AGN.  For narrow-line AGN, the point source is typically much dimmer than the host galaxy, allowing one to ignore its effects on total optical flux and color for statistical samples.  Despite its relative faintness, however, the presence of a point source can substantially alter the shape of the surface brightness distribution.  This leads to morphological measurements biased toward higher light concentrations and lower asymmetry.

Selection effects derive from the manifold methods for identifying AGN, each of which reveals a different facet of AGN activity.  The properties of the host galaxies of AGN vary with luminosity, black hole mass, and probably redshift, so different samples are difficult to compare.  Where early quasar host galaxy studies picked out the brightest quasars living in the most massive galaxies, more recent studies employ broad-band optical colors, optical spectra, X-ray point source luminosities, or infrared colors to select AGN.  Furthermore, the studies mentioned above encompass the full range of redshift-areal coverage space, from large numbers of low-redshift AGN observed in the SDSS, to modest numbers of higher-redshift AGN observed in the pencil-beam surveys like GEMS, to the handful of high-redshift AGN hosts observed through strong lenses.  COSMOS falls into the pencil-beam category, yielding a substantial sample size at moderate redshifts.  Although the X-ray luminosity range ($42\lesssim \log L_x \lesssim 44$) of our X-ray sample spans moderate-to-powerful AGN, the bolometric luminosities of our radio AGN are difficult to estimate and are likely lower.  This makes comparisons even within COSMOS a challenge.

We can attempt to examine biases associated with luminosity here.  Figure \ref{fig.sersic_vs_xlum} shows the measured host galaxy Sersic index versus X-ray luminosity for X-ray AGN.  We find no notable trend, suggesting that earlier studies suffered little selection bias, although optical quasar selection is not directly comparable to X-ray selection.  The lack of a trend of host morphology with X-ray luminosity might also have physical implications for the merger picture mentioned above.  As the gaseous galaxies pass by each other then converge, one might naively expect the highest X-ray luminosities to coincide with final coalescence into a galactic bulge.  Our data do not support such a scenario, although we must defer analysis of the details to theoretical work.

We also estimate the host galaxy luminosities based on our measurements of the apparent magnitude.  To compute the rest-frame $V$-band magnitudes, $M_V$, we use the spectroscopic redshifts and assume spectral energy distributions for the AGN host galaxies.  We choose the rest-frame $V$-band because it shifts into the observed $I$-band near the median redshift of our sample, and because it serves as a convenient reference to the absolute $V$-band magnitudes derived for all galaxies in the COSMOS photometric redshift catalog.  Following \citet{hog02}, we calculate the K-corrections by applying filter curves for the $F814W$ filter of HST and the Subaru $V$ filter used for COSMOS observations.  We calculate the corrections with both an elliptical galaxy and an Sb galaxy template optical SED from \citet{kin96}, and we display results from the early-type template.  Given that AGN host galaxies have blue colors and recent star formation, the true SED lies somewhere between the two templates considered here.  However, at all redshifts considered here, the K-correction differs by $\lesssim$0.2 magnitudes between the two templates, so the choice of template does not strongly affect the results.  Figure \ref{fig.sersic_vs_hostlum} shows measured Sersic index versus these derived host galaxy absolute magnitudes.  We note that for both X-ray AGN and radio AGN hosts, the distribution of absolute magnitudes peaks around $M_V=-22$, so these galaxies have similar luminosities to $M_V^*\simeq -22$ (for $z=1$, computed by starting from the local value of \citealt{bro01} and following \citealt{cap07} and \citealt{smi05} in allowing 1 magnitude of passive evolution to $z=1$).  We see a weak trend (with correlation coefficient $\approx -0.2$ for X-ray and $\approx -0.4$ for radio AGN) of morphology with host galaxy luminosity in both samples, where brighter host galaxies have bulge-dominated morphologies.  This trend is not redshift-dependent, and may reflect the general galaxy population.

Although the major merger picture is elegant and enticing, none of our AGN samples shows enhancement of the merger and interaction indicators applied.  This roughly agrees with previous studies \citep{gro05, pie07}.  Furthermore, differences in this result between subsamples are not significant: radio and X-ray AGN candidates all follow the same trends as non-active galaxies.  These results suggest that major galaxy mergers do not play the dominant role in triggering AGN activity, with the likely alternatives being minor mergers and interactions, and dynamical instabilities within galaxies \citep[cf/]{has08}.

We caution, however, that the tools we apply here may be too blunt to cut to the heart of the question.  The key uncertainty in drawing conclusions from tests like these is the timescale -- for any given merger event, how long it takes to go from interaction to merger to coalescence to relaxation, when and how long an AGN fueling event might occur, and how long interaction indicators will be observable.  While galaxy counts in AGN environments may be connected to the likelihood for mergers, they serve as an indirect probe at best.  By counting neighbors we are finding systems that are likely to merge in the future, rather than those that have already merged and might be in the midst of AGN fueling.

Like galaxy counts, the morphological measures used in this work may not trace galaxy mergers as sensitively as necessary to distinguish recently merged systems from normal galaxies at moderate redshifts.  Certainly we can be confident that high-$A$ galaxies are undergoing mergers, but not all recent mergers necessarily have large values of asymmetry.  As \citet{con03} suggests, perhaps only systems in certain phases of the merger process exhibit the large-scale asymmetries to which $A$ is sensitive.  If major mergers are indeed the trigger of AGN, the triggering lags the merger in such a way that the dominant light distribution of the host galaxy appears essentially as relaxed as a normal galaxy.  This presents difficulties becauses the typical AGN duty cycle of $10^8$ years is much shorter than a typical galaxy's relaxation time, and similar to the free-fall time on which violent relaxation is expected to occur.  Another possibility is that large scale disruptions such as tidal tails are present, but at such a low surface brightness that they are too difficult to see and measure with such crude techniques at higher redshifts.

Perhaps minor mergers are the answer. These interactions may disrupt a gaseous galaxy enough to cause central inflow of gas onto a black hole without significantly altering the observed distribution of light.  As emphasized in the review by \citet{jog06}, for most AGN the amount of fuel available is not the problem per se, but rather decreasing the specific angular momentum of that fuel by 99.99\% to feed the central black hole.  Minor mergers could potentially disrupt the inner regions of a gas-rich galaxy enough to initiate fueling, although this is not seen in simulations.  Such an interaction would be difficult to detect using the techniques described here.  However, \citet{can07} find faint shell structures indicative of a merger in deep HST images of the host galaxy of a low-redshift quasar whose morphology had previously been considered quiescent.  Detecting similar structures in larger samples of AGN could help reveal a merger-driven fueling mechanism, but such work will be challenging at moderate and high redshifts.

These considerations make clear the need for more work in understanding the relationships among dynamics, timescales, and observable properties of galaxy mergers and AGN.  As the resolution and scale of simulations improves, we expect new constraints on the merger mass ratios necessary to trigger black hole accretion to emerge.  In future work, we hope to develop new techniques for detecting and measuring low surface brightness features that may betray recent merger activity.

\section{Summary and Conclusion}
\label{sec.conclusion}
We explored the host morphologies and environments of AGN in the COSMOS field.  Using X-ray- and radio-selected AGN candidates with confirmed spectroscopic redshifts, we analyzed host galaxy structural properties as well as merger indicators, probing the connection between AGN activity and galaxy interactions.  The following summarizes our main points: 

\begin{itemize}
\item The central point source in optical images of X-ray selected AGN has substantial impact on measured structural parameters such as asymmetry and concentration.  Insufficient accounting for the point source can lead to systematically low asymmetry and systematically high concentration.
\item Full 2-D fits and concentration measurements which account for the central point source in X-ray AGN show that their host galaxies have a broad range of morphologies whose distribution is intermediate between the bulge- and disk-dominated regimes.
\item Although radio AGN hosts also appear to have a wide range of morphologies, contamination by star-forming galaxies prevents us from distinguishing them from early-type normal galaxies.
\item Measurements of AGN host galaxy asymmetry do not differ significantly from those of matched control galaxies.
\item Neighbor counts around AGN are indistinguishable from those around matched control galaxies using photometric redshifts.
\end{itemize}

These findings do not support the hypothesis that major mergers drive black hole activity, but they do suggest that the host galaxies of AGN at these luminosities may be in a state of morphological transition.  Future work by members of the COSMOS collaboration will address this possibility in more detail by examing the colors and star formation rates of AGN hosts and their relationships with environment (Silverman et al. in preparation).

 
 \acknowledgments

We gratefully acknowledge the contributions of the entire COSMOS colaboration
 consisting of more than 70 scientists. More information on the COSMOS survey is available at
  {\bf \url{http://www.astro.caltech.edu/cosmos}}.  We also thank the anonymous, thorough referee for contributing to the quality of this work.  JG would like to thank Romeel Dave, Rodger Thompson, and Ann Zabludoff for guidance and useful discussion, and St\'{e}phanie Juneau for help with K-corrections.  JDR was supported by internal funding at the Jet Propulsion Laboratory, California Institute of Technology, operated under a contract with NASA.  KJ acknowledges support by the German DFG under grant SCHI~536/3-1 andthrough the DFG Emmy Noether-Program under grant JA~1114/3-1.
 
 
 
 {\it Facilities:} \facility{HST (ACS)}, \facility{Magellan:Baade (IMACS)}, \facility{XMM ()}, \facility{VLA ()}, \facility{MMT (Hectospec)}.

 \clearpage
\clearpage

\clearpage
 \pagestyle{empty}
\begin{landscape}
\begin{deluxetable}{ccccccccccccccccc}
\tablecaption{Results of 2-D fits. \label{tab.fits}}
\tabletypesize{\scriptsize}
\tablewidth{0pt}
\tablehead{
\colhead{Obj} 
& \colhead{RA \tablenotemark{a}} 
& \colhead{Dec \tablenotemark{a}} 
& \colhead{z \tablenotemark{b}} 
& \colhead{$m_h$ \tablenotemark{c,d}} 
& \colhead{$m_h^*$ \tablenotemark{c}} 
& \colhead{$r_h$ \tablenotemark{c,d}} 
& \colhead{$r_h^*$ \tablenotemark{c}} 
& \colhead{$n$ \tablenotemark{c,d}} 
& \colhead{$n^*$ \tablenotemark{c}} 
& \colhead{$m_p$ \tablenotemark{c,d}} 
& \colhead{$A$ \tablenotemark{e}} 
& \colhead{$C$ \tablenotemark{e}} 
& \colhead{$r_{\eta_0}$ \tablenotemark{e}} 
& \colhead{flag \tablenotemark{f}} 
& \colhead{class \tablenotemark{g}} 
\\     & \colhead{(degrees)} & \colhead{(degrees)} & 
& \colhead{(F184W)} & \colhead{(F814W)} & \colhead{(pixels)} & \colhead{(pixels)} & & & \colhead{(F814W)} & & & & \colhead{(pixels)} & &
}
\startdata
COSMOS$\_$J100045.16+024133.1 & 150.1882 & 2.6925 & 0.30 & 21.06(0.60) & 21.06 &  40.07( 30.29) &  40.07 & 5.14(1.30) & 5.14 & 26.87( 0.04) & 0.70 &  2.47 &  68.3 & 3 & r23  \\
COSMOS$\_$J095837.34+013710.4 & 149.6556 & 1.6196 & 0.30 & 19.96(0.35) & 19.94 &  24.97(  5.03) &  25.19 & 2.41(0.48) & 2.50 & 25.98( 0.02) & 0.20 &  2.04 &  50.9 & 0 & r23  \\
COSMOS$\_$J095950.46+021310.1 & 149.9603 & 2.2195 & 0.31 & 19.08(0.31) & 18.62 &  64.42( 34.51) & 142.30 & 2.58(0.67) & 8.00 & 22.17( 0.05) & 0.28 &  1.50 &  81.3 & 0 & x3  \\
COSMOS$\_$J095944.46+020858.7 & 149.9353 & 2.1496 & 0.31 & 19.21(0.15) & 19.10 &  35.40(  5.83) &  38.53 & 1.18(0.21) & 1.61 & 23.65( 0.01) & 0.16 &  1.57 &  64.7 & 0 & r23  \\
COSMOS$\_$J095808.99+014131.2 & 149.5375 & 1.6920 & 0.31 & 19.63(0.27) & 19.60 &  48.88( 34.48) &  48.92 & 0.75(0.43) & 0.84 & 24.96( 0.01) & 0.59 &  1.40 &  83.3 & 0 & r2  \\
COSMOS$\_$J095832.12+020656.9 & 149.6338 & 2.1158 & 0.31 & 17.27(0.35) & 17.32 & 141.20( 56.79) & 127.90 & 6.14(0.92) & 5.99 & 24.27( 0.09) & 0.35 &  2.03 & 129.9 & 0 & r3  \\
COSMOS$\_$J100218.75+015815.5 & 150.5781 & 1.9710 & 0.31 & 20.35(0.08) & 20.09 &   9.36(  0.79) &  10.02 & 1.59(0.11) & 3.58 & 22.71( 0.01) & 0.10 &  1.61 &  22.1 & 0 & r2  \\
COSMOS$\_$J095834.95+015348.4 & 149.6456 & 1.8968 & 0.31 & 20.84(0.15) & 20.82 &  20.00(  3.68) &  19.90 & 0.66(0.22) & 0.70 & 25.45( 0.01) & 0.19 &  1.48 &  38.8 & 0 & r23  \\
COSMOS$\_$J100042.08+022534.2 & 150.1754 & 2.4262 & 0.31 & 19.35(0.36) & 19.43 &  24.67(  3.70) &  20.02 & 2.79(0.47) & 2.32 & 23.09( 0.02) & 1.13 &  1.25 &  55.9 & 1 & r2  \\
COSMOS$\_$J095842.01+015442.4 & 149.6750 & 1.9118 & 0.31 & 21.45(0.63) & 21.54 &  45.52( 52.47) &  36.49 & 2.35(1.12) & 2.10 & 25.82( 0.01) & 0.87 &  2.24 &  96.9 & 0 & x2  \\
COSMOS$\_$J095846.75+023910.8 & 149.6948 & 2.6530 & 0.32 & 20.81(0.28) & 20.64 &  30.87( 20.88) &  34.00 & 1.14(0.43) & 1.44 & 25.21( 0.01) & 0.63 &  1.77 &  83.6 & 1 & x2  \\
COSMOS$\_$J095837.92+024708.2 & 149.6580 & 2.7856 & 0.32 & 20.10(0.28) & 20.05 &  42.08( 28.37) &  43.22 & 1.18(0.44) & 1.43 & 25.06( 0.01) & 0.38 &  1.83 &  79.5 & 0 & r2  \\
COSMOS$\_$J095749.02+015310.1 & 149.4543 & 1.8861 & 0.32 & 20.29(0.47) & 20.18 &  16.13(  2.77) &   6.84 & 3.51(0.62) & 8.00 & 21.81( 0.03) & 0.19 &  1.95 &  36.3 & 0 & x23  \\
COSMOS$\_$J095806.72+020738.1 & 149.5280 & 2.1273 & 0.32 & 20.79(0.22) & 20.79 &  23.67(  6.71) &  23.67 & 2.38(0.52) & 2.38 & 24.27( 0.02) & 0.61 &  1.33 &  48.7 & 3 & r3  \\
COSMOS$\_$J095845.62+014016.1 & 149.6901 & 1.6711 & 0.32 & 18.53(0.43) & 18.47 & 108.60(110.61) & 114.30 & 2.08(0.76) & 2.43 & 23.76( 0.04) & 0.47 &  1.92 & 144.2 & 0 & r3  \\
\enddata
\tablenotetext{a}{Positions in Right Ascension and Declination are for optical counterparts to the XMM-Newton X-ray point-source catalog \citep{cappell07, bru07} and the VLA radio source catalog \citep{sch07}.}
\tablenotetext{b}{Redshifts are derived from optical spectra from Magellan/IMACS and MMT/Hectospec \citep{tru07}.}
\tablenotetext{c}{Best-fit morphological parameters based on 2-D surface brightness fits are: host galaxy apparent F814W magnitude ($m_h$), host galaxy effective radius in pixels ($r_h$; 1 pixel $=$ 0.03 arcseconds), host galaxy Sersic index ($n$), and nuclear point source F814W magnitude ($m_p$).  Asterisks denote best-fit parameters for fits which do not include a nuclear point source componenent.}
\tablenotetext{d}{Uncertainty estimates for the best-fit parameters (shown in parentheses) are based on fits to our simulated AGN images, and give the 1-$\sigma$ scatter on each parameter for an AGN with the given mean surface brightness.}
\tablenotetext{e}{Non-parametric morphological indicators are: Asymmetry ($A$), Concentration ($C$), and Petrosian radius ($r_{\eta_0}$).}
\tablenotetext{f}{fit flag -- 0:Successful automated fit, 1:fit runs into boundaries, 2:Successful manual fit, 3:Fit better without PSF component, 4:Poor fit (subjective)}
\tablenotetext{g}{Spectral classes from \citet{tru07}.  The first letter indicates X-ray (``x'') or radio (``r'') selected objects, and the numbers indicate optical spectral types -- 1:Broad line AGN, 2:Narrow lines, 23:Hybrid Narrow line/Red galaxy, 3:Red Galaxy}
\end{deluxetable}
\clearpage
\end{landscape}

\begin{deluxetable}{ccccccccc}
\tablecaption{Results of PSF-subtraction for broad-line AGN. \label{tab.type1}}
\tabletypesize{\tiny}
\tablewidth{0pt}
\tablehead{
\colhead{Obj} 
& \colhead{RA \tablenotemark{a}} 
& \colhead{Dec \tablenotemark{a}} 
& \colhead{z \tablenotemark{b}} 
& \colhead{$m_h$ upper limit \tablenotemark{c}} 
& \colhead{$m_h$ best fit \tablenotemark{d}} 
& \colhead{$m_p$ lower limit \tablenotemark{c}} 
& \colhead{$m_p$ best fit \tablenotemark{d}} 
& \colhead{flag \tablenotemark{e}} 
\\ & \colhead{degrees} & \colhead{degrees} & & & & & &
}
\startdata
COSMOS$\_$J095902.76+021906.4 & 149.7615 & 2.3185 & 0.34 & 19.5 & 18.41 & 19.1 & 19.57 & 0  \\
COSMOS$\_$J095928.32+022106.9 & 149.8680 & 2.3519 & 0.35 & 20.4 & 19.27 & 21.5 & 21.65 & 0  \\
COSMOS$\_$J100043.15+020637.2 & 150.1798 & 2.1103 & 0.36 & 19.2 & 18.15 & 20.3 & 20.40 & 1  \\
COSMOS$\_$J100212.11+014232.4 & 150.5505 & 1.7090 & 0.37 & 21.1 & 20.33 & 20.7 & 20.97 & 4  \\
COSMOS$\_$J100025.25+015852.3 & 150.1052 & 1.9812 & 0.37 & 20.9 & 18.62 & 18.9 & 28.88 & 4  \\
COSMOS$\_$J100243.96+023428.6 & 150.6832 & 2.5746 & 0.38 & 20.0 & 18.80 & 20.2 & 20.47 & 0  \\
COSMOS$\_$J095909.54+021916.5 & 149.7897 & 2.3213 & 0.38 & 20.8 & 20.15 & 20.9 & 21.05 & 0  \\
COSMOS$\_$J100033.49+013811.6 & 150.1395 & 1.6366 & 0.52 & 22.0 & 21.28 & 21.5 & 22.07 & 0  \\
COSMOS$\_$J100118.53+015543.0 & 150.3272 & 1.9286 & 0.53 & 20.8 & 20.48 & 21.7 & 21.84 & 0  \\
COSMOS$\_$J100046.73+020404.5 & 150.1947 & 2.0679 & 0.55 & 20.7 & 19.97 & 20.7 & 20.76 & 0  \\
COSMOS$\_$J100141.10+021260.0 & 150.4212 & 2.2167 & 0.62 & 22.3 & 21.94 & 22.1 & 22.38 & 0  \\
COSMOS$\_$J100230.06+014810.4 & 150.6252 & 1.8029 & 0.63 & 21.3 & 20.78 & 19.7 & 19.90 & 0  \\
COSMOS$\_$J095938.99+021201.3 & 149.9124 & 2.2004 & 0.69 & 21.6 & 21.34 & 20.7 & 20.79 & 0  \\
COSMOS$\_$J100012.91+023522.8 & 150.0538 & 2.5897 & 0.70 & 20.6 & 19.25 & 19.0 & 20.13 & 4  \\
COSMOS$\_$J095813.33+020536.2 & 149.5555 & 2.0934 & 0.70 & 21.2 & 20.58 & 21.4 & 21.57 & 0  \\
COSMOS$\_$J095817.54+021938.7 & 149.5731 & 2.3274 & 0.73 & 22.9 & 22.88 & 21.2 & 21.30 & 4  \\
COSMOS$\_$J095938.55+023316.9 & 149.9106 & 2.5547 & 0.75 & 20.8 & 19.96 & 20.7 & 24.67 & 4  \\
COSMOS$\_$J100202.22+024157.8 & 150.5093 & 2.6994 & 0.79 & 22.2 & 22.12 & 21.4 & 21.51 & 0  \\
COSMOS$\_$J100129.83+023239.0 & 150.3743 & 2.5442 & 0.83 & 21.6 & 20.08 & 21.6 & 21.98 & 4  \\
COSMOS$\_$J100003.27+014802.2 & 150.0136 & 1.8006 & 0.83 & 22.2 & 21.85 & 22.7 & 23.05 & 2  \\
COSMOS$\_$J100033.38+015237.2 & 150.1391 & 1.8770 & 0.83 & 21.6 & 20.81 & 20.8 & 27.15 & 4  \\
COSMOS$\_$J095809.93+021057.7 & 149.5414 & 2.1827 & 0.84 & 22.2 & 22.18 & 21.8 & 21.89 & 4  \\
COSMOS$\_$J100002.21+021631.8 & 150.0092 & 2.2755 & 0.85 & 21.4 & 19.29 & 21.0 & 21.06 & 4  \\
COSMOS$\_$J100159.43+023935.6 & 150.4976 & 2.6599 & 0.85 & 21.4 & 21.02 & 21.0 & 21.00 & 0  \\
COSMOS$\_$J100229.33+014528.1 & 150.6222 & 1.7578 & 0.88 & 22.2 & 22.43 & 20.2 & 20.20 & 4  \\
COSMOS$\_$J100147.90+021447.2 & 150.4496 & 2.2465 & 0.88 & 21.9 & 20.83 & 20.7 & 21.42 & 4  \\
COSMOS$\_$J100120.25+020341.2 & 150.3344 & 2.0614 & 0.91 & 21.8 & 21.31 & 20.6 & 20.77 & 0  \\
COSMOS$\_$J095946.92+022209.5 & 149.9455 & 2.3693 & 0.91 & 22.8 & 22.18 & 21.1 & 21.11 & 0  \\
COSMOS$\_$J100055.62+013954.9 & 150.2318 & 1.6652 & 0.91 & 22.7 & 21.67 & 22.6 & 23.77 & 4  \\
COSMOS$\_$J100116.28+023607.5 & 150.3178 & 2.6021 & 0.96 & 21.7 & 21.35 & 21.2 & 21.21 & 0  \\
COSMOS$\_$J100151.11+020032.7 & 150.4630 & 2.0091 & 0.96 & 22.0 & 20.27 & 20.1 & 20.84 & 4  \\
COSMOS$\_$J100141.33+021031.5 & 150.4222 & 2.1754 & 0.98 & 21.8 & 21.06 & 20.9 & 21.65 & 4  \\
COSMOS$\_$J100202.78+022434.6 & 150.5116 & 2.4096 & 0.99 & 22.1 & 19.98 & 20.6 & 30.57 & 4  \\
COSMOS$\_$J100114.86+020208.8 & 150.3119 & 2.0358 & 0.99 & 21.4 & 20.55 & 22.1 & 22.34 & 0  \\
\enddata
\tablenotetext{a}{Positions in Right Ascension and Declination are for optical counterparts to the XMM-Newton X-ray point-source catalog \citep{cappell07, bru07} and the VLA radio source catalog \citep{sch07}.}
\tablenotetext{b}{Redshifts are derived from optical spectra from Magellan/IMACS and MMT/Hectospec \citep{tru07}.}
\tablenotetext{c}{Upper limit on host galaxy F814W apparent magnitude ($m_h$), and lower limit on nuclear point source apparent magnitude ($m_p$), are based on PSF-only fit and subtraction.}
\tablenotetext{d}{Best fit apparent magnitudes are based on 2-D surface brightness fitting.}
\tablenotetext{e}{Flag for goodness of fit.  See note f of Table \ref{tab.fits}.}
\end{deluxetable}

\clearpage

 

\begin{thebibliography}{}

\bibitem[Abraham \etal (1994)]{abr94} Abraham, R. G., Valdes, F., Yee, H. K. C., \& van den Bergh, S. 1994, ApJ, 432, 75.

\bibitem[Abraham \etal (1996)]{abr96} Abraham, R. G., van den Bergh, S., Glazebrook, K., Ellis, R., Santiago, B. X., Surma, P., \& Griffiths, R. E. 1996, ApJS, 107, 1.

\bibitem[Alonso \etal (2007)]{alo07} Alonso, M. S., Lambas, D. G., Tissera, P., \& Coldwell, G. 2007, MNRAS, 375, 1017.

\bibitem[Baldwin \etal (1981)]{bal81} Baldwin, J. A., Phillips, M. M., \& Terlevich, R. 1981, PASP, 93, 5.

\bibitem[Bauer \etal (2004)]{bau04} Bauer, F. E., Alexander, D. M., Brandt, W. N., Schneider, D. P., Treister, E., Hornschemeier, A. E., Garmire, G.P.  2004, AJ, 128, 2048.

\bibitem[Bertin \& Arnouts (1996)]{ber96} Bertin, E. and Arnouts, S. 1996.  A\&AS, 117, 393.

\bibitem[Brown \etal (2001)]{bro01} Brown, W. R., Geller, M. J., Fabricant, D. G., Kurtz, M. J. 2001, AJ 122, 714.

\bibitem[Brusa \etal (2007)]{bru07} Brusa, M., \etal 2007. ApJS, 172, 353.

\bibitem[Bundy \etal (2007)]{bun07} Bundy, K. \etal~ 2007, ApJ, submitted (arXiv:0710.2105).

\bibitem[Canalizo \etal (2007)]{can07} Canalizo, G., Bennert, N., Jungwiert, B., Stockton, A., Schweizer, F., Lacy, M., \& Peng, C. 2007, ApJ, 669, 801.

\bibitem[Capak \etal (2007)]{cap07} Capak, P., Abraham, R. G., Ellis, R. S., Mobasher, B., Scoville, N., Sheth, K., \& Koekemoer, A. 2007, ApJS, 172, 284.

\bibitem[Capak \etal (2007)]{cap07b} Capak, P. \etal 2007b, ApJS, 172, 99.

\bibitem[Cappelluti \etal (2007)]{cappell07} Cappelluti, N. \etal 2007, ApJS, 172, 341.

\bibitem[Coil \etal (2007)]{coi07} Coil, A. L., Hennawi, J. F., Newman, J. A., Cooper, M. C., \& Davis, M. 2007, ApJ, 654, 115.

\bibitem[Condon \etal (1998)]{con98} Condon, J. J., Cotton, W. D., Greisen, E. W., Yin, Q. F., Perley, R. A., Taylor, G. B., \& Broderick, J. J.  1998, AJ, 115, 1693.

\bibitem[Conselice (2000)]{con00} Conselice, C. J., Bershady, M. A., \& Jangren, A. 2000, ApJ, 529, 886.

\bibitem[Conselice (2003)]{con03} Conselice, C. J. 2003, ApJS, 147, 1.

\bibitem[De Robertis \etal (1998)]{der98} De Robertis, M. M., Yee, H. K. C., \& Hayhoe, K. 1998.  ApJ, 496, 93D.

\bibitem[de Vaucouleurs \& Capaccioli (1979)]{dev79} de Vaucouleurs, G. \& Capaccioli, M. 1979.  ApJS, 40, 699.

\bibitem[Dunlop \etal (2003)]{dun03} Dunlop, J. S., McLure, R. J., Kukula, M. J., Baum, S. A., O'Dea, C. P., \& Hughes, D. H. 2003, MNRAS, 340, 1095.

\bibitem[Ferrarese (2004)]{fer04} Ferrarese, L. 2004, in Supermassive Black Holes in the Distant Universe, ed. Barger, A. (Netherlands: Kluwer Academic Publishers), 1.

\bibitem[Ferrarese \& Merritt(2000)]{fer00} Ferrarese, L. \& Merritt, D. 2000.  ApJ, 539, L9.

\bibitem[Gebhardt \etal (2000)]{geb00} Gebhardt, K., \etal~ 2000, ApJ, 539, 13.

\bibitem[Grogin \etal (2005)]{gro05} Grogin, N. A. \etal~ 2005, ApJ, 627L, 97.

\bibitem[Hasinger \etal (2007)]{has07} Hasinger, G. \etal~ 2007, ApJS, 172, 29.

\bibitem[Hasinger (2008)]{has08} Hasinger, G. 2008, A\&A, submitted.

\bibitem[H\"au{\ss}ler \etal (2007)]{hau07} H\"au{\ss}ler, B. \etal~ 2007, ApJS, 172, 615.

\bibitem[Hogg \etal (2002)]{hog02} Hogg, D. W., Baldry, I. K., Blanton, M. R., Eisenstein, D. J. 2002, preprint (astro-ph/0210394)

\bibitem[Hopkins \etal (2006)]{hop06} Hopkins, P. F., Hernquist, L., Cox, T. J., Robertson, B., Springel, V. 2006, ApJS, 163, 50.

\bibitem[Hopkins \& Hernquist (2006)]{hop06b} Hopkins, P. F. \& Hernquist, L. 2006.  ApJS, 166, 1H.

\bibitem[Jahnke \etal (2004a)]{jah04a} Jahnke, K., Kuhlbrodt, B., \& Wisotzki, L. 2004, MNRAS, 352, 399.

\bibitem[Jahnke \etal (2004b)]{jah04} Jahnke, K. \etal~ 2004.  ApJ, 614, 568.

\bibitem[Jogee (2006)]{jog06} Jogee, S. 2006.  LNP, 693, 143.

\bibitem[Kauffmann \etal (2003)]{kau03} Kauffmann, G. \etal~ 2003, MNRAS, 346, 1055.

\bibitem[Kewley \etal (2001)]{kew01} Kewley, L. J., Dopita, M. A., Sutherland, R. S., Heisler, C. A., \& Trevena, J. 2001, ApJ, 556, 121.

\bibitem[Kim \etal (2008)]{kim08} Kim, M., Ho, L. C., Peng, C. Y., Barth, A. J., \& Im, M. 2008, ApJS, in press (arXiv:0807.1334)

\bibitem[Kinney \etal (1996)]{kin96} Kinney, A. L., Calzetti, D., Bohlin, R. C., McQuade, K., Storchi-Bergmann, T., \& Schmitt, H. R. 1996, ApJ, 467, 38.

\bibitem[Koekemoer \etal (2007)]{koe07} Koekemoer, A. M. \etal~ 2007, ApJS, 172, 196.

\bibitem[Kormendy \& Richstone (1995)]{kor95} Kormendy, J. \& Richstone, D. 1995, ARA\&A, 581.

\bibitem[Krist (2003)]{kri03} Krist, J. Instrument Science Rep. ACS 2003-06 (Baltimore: STScI)

\bibitem[Lilly \etal (2007)]{lil07} Lilly, S. J. \etal 2007, ApJS, 172, 70.

\bibitem[Magorrian \etal (1998)]{mag98} Magorrian, J., \etal~ 1998, AJ, 115, 2285.

\bibitem[Marconi \& Hunt (2003)]{mar03} Marconi, A. \& Hunt, L.K. 2003, ApJ, 589, L21.

\bibitem[McLure \& Dunlop (2001)]{mcl01} McLure, R.J. \& Dunlop, J.S. 2001, MNRAS, 327, 199.

\bibitem[McLeod \& McLeod (2001)]{mm01} McLeod, K. K. \& McLeod, B. A. 2001, ApJ, 546, 782.

\bibitem[Menci \etal (2004)]{men04} Menci, N., Cavaliere, A., Fontana, A., Giallongo, E., Poli, F., \& Vittorini, V. 2004, ApJ, 604, 12M.

\bibitem[Mobasher \etal (2007)]{mob07} Mobasher, B. \etal~ 2007, ApJS, 172, 117.

\bibitem[Mushotzky (2004)]{mus04} Mushotzky, R. 2004, in Supermassive Black Holes in the Distant Universe, ed. Barger, A. (Netherlands: Kluwer Academic Publishers), 53.

\bibitem[Odell \etal (2002)]{ode02} Odell, A. P., Schombert, J., \& Rakos, K.  2002, AJ, 124, 3061.

\bibitem[Patton \etal (2000)]{pat00} Patton, D. R., Carlberge, R. G., Marzke, R. O., Pritchet, C. J., da Costa, L. N., \& Pellegrini, P. S. 2000, ApJ, 536, 153.

\bibitem[Patton \etal (2002)]{pat02} Patton, D. R. \etal~ 2002.  ApJ, 565, 208.

\bibitem[Peng \etal (2002)]{pen02} Peng, C. Y., Ho, L. C., Impey, C. D., \& Rix, H-W. 2002, AJ, 124, 266.

\bibitem[Peng \etal (2006a)]{pen06a} Peng, C. Y., Impey, C. D., Ho, L. C., Barton, E. J., \& Rix, H-W. 2006a, ApJ, 640, 114.

\bibitem[Peng \etal (2006b)]{pen06b} Peng, C. Y., Impey, C. D., Rix, H-W., Kochanek, C. S., Keeton, C. R., Falco, E. E., Lehar, J., \& McLeod, B. A., 2006b, ApJ, 649, 616.

\bibitem[Peng \etal (2006c)]{pen06c} Peng, C. Y., Impey, C. D., Rix, H-W., Falco, E. E., Keeton, C. R., Kochanek, C. S., Lehar, J., \& McLeod, B. A. 2006c, NewAR, 50, 689.

\bibitem[Petrosian (1976)]{pet76} Petrosian, V., 1976.  ApJ, 209, L1.

\bibitem[Pierce \etal (2007)]{pie07} Pierce, C. M. \etal~ 2007, ApJL, 660L, 19.

%
\bibitem[Rhodes \etal (2007)]{rho07} Rhodes, J. \etal~ 2007.  ApJS, 172, 203.

\bibitem[Sanchez \etal (2004)]{san04} Sanchez, S. F. \etal~ 2004.  ApJ, 614, 586.
    
\bibitem[Schinnerer \etal (2007)]{sch07} Schinnerer, E. \etal~ 2007, ApJS, 172, 46.

\bibitem[Scoville \etal (2007)]{sco07a} Scoville, N. \etal~ 2007, ApJS, 172, 1.

\bibitem[Scoville \etal (2007)]{sco07b} Scoville, N. \etal~ 2007, ApJS, 172, 150.

\bibitem[Serber \etal (2006)]{ser06} Serber, W., Bahcall, N., Menard, B., \& Richards, G. 2006, ApJ, 643, 68.

\bibitem[Sersic (1968)]{ser68} Sersic, J. L.  Atlas de Galaxias Australes.  (Cordoba: Obbs. Astron. Univ. Nacional Cordoba).

\bibitem[Silverman \etal (2008)]{sil08} Silverman, J. D. \etal~ 2007, ApJ, 675, 1025.

\bibitem[Smith \etal (2005)]{smi05} Smith, G. P., Treu, T., Ellis, R. S., Moran, S. M., \& Dressler, A. 2005, ApJ, 620, 78.

\bibitem[Smol\v{c}i\'{c} \etal (2006)]{smo06} Smol\v{c}i\'{c}, V. \etal~ 2006, MNRAS, 371, 121.

\bibitem[Smol\v{c}i\'{c} \etal (2008)]{smo08} Smol\v{c}i\'{c}, V. \etal~ 2008, ApJS, in press (arXiv:0803.0997)

\bibitem[Springel \etal (2005)]{spr05} Springel, V., Di Matteo, T., \& Hernquist, L. 2005, MNRAS, 361, 776.

\bibitem[Strand \etal (2008)]{str08} Strand, N. E., Brunner, R. J., \& Myers, A. D. 2008, preprint (arXiv:0712.2474).

\bibitem[Tremonti \etal (2007)]{tre07} Tremonti, C. A., Moustaka, J., \& Diamond-Stanic, A. M. 2007, ApJ, 663, 77.

\bibitem[Treu \etal (2004)]{tre04} Treu, T., Malkan, M. A., \& Blandford, R. D. 2004, ApJ, 615L, 97.

\bibitem[Trump \etal (2007)]{tru07} Trump, J. R. \etal~ 2007, ApJS, 172, 383.

\bibitem[Veilleux \& Osterbrock (1987)]{vei87} Veilleux, S. \& Osterbrock, D. E. 1987.  ApJS, 63, 295.

\end{thebibliography}
 \end{document}